\documentclass[11pt,a4paper]{article}
\pdfoutput=1

\usepackage{jheppub}
\usepackage{latexsym}
\usepackage{multirow}
\usepackage{color}
\usepackage[usenames,dvipsnames,table]{xcolor}
\usepackage{graphicx}
\usepackage{epsfig}  
\usepackage{epsf}    
\usepackage{dcolumn}
\usepackage{bm}
\usepackage{dcolumn}
\usepackage{textcomp}
\usepackage{float}
\usepackage{subfig}
\usepackage{hypcap}
\usepackage[]{hyperref}
\usepackage{makecell}
\usepackage{color}
\usepackage{pifont}
\usepackage{appendix}
\usepackage{amsmath}
\usepackage{multirow,bigdelim}  
\usepackage{lineno}
\usepackage[normalem]{ulem}
\usepackage{adjustbox}

\hypersetup{
  bookmarks=true,         
  unicode=false,          
  pdftoolbar=true,        
 pdfmenubar=true,        
 pdffitwindow=true,     
 pdfstartview={FitH},    
 pdfsubject={Neutrino Oscillations Phenomenology},   
 pdfnewwindow=true,      
 pdfcreator={RevTeX},
 colorlinks=true,       
 linkcolor=red,          
 citecolor=blue,        
 filecolor=black,      
 urlcolor=blue,           
  }

\newcommand{\aee}{\ensuremath{\alpha_{11}}}
\newcommand{\ame}{\ensuremath{\alpha_{21}}}
\newcommand{\amm}{\ensuremath{\alpha_{22}}}
\newcommand{\ate}{\ensuremath{\alpha_{31}}}
\newcommand{\atm}{\ensuremath{\alpha_{32}}}
\newcommand{\att}{\ensuremath{\alpha_{33}}}
\newcommand{\ie}{\textit{i.e.}}
\newcommand{\nue}{\mbox{$\nu_e$}}

\newcommand{\capdef}{}
\newcommand{\mycaption}[2][\capdef]{\renewcommand{\capdef}{#2}
	\caption[#1]{{\footnotesize #2}}}

\preprint{IP/BBSR/2021-10}

\title{Model-Independent Constraints on Non-Unitary Neutrino Mixing 
from High-Precision Long-Baseline Experiments}

\author[a,b,c,d]{Sanjib Kumar Agarwalla,}
\author[a,b]{Sudipta Das,}
\author[e]{Alessio Giarnetti,}
\author[e]{Davide Meloni} 

\affiliation[a]{Institute of Physics, Sachivalaya Marg, Sainik School Post, Bhubaneswar 751005, India}
\affiliation[b]{Homi Bhabha National Institute, Training School Complex, Anushakti Nagar, Mumbai 400094, India}
\affiliation[c]{International Centre for Theoretical Physics, Strada Costiera 11, 34151 Trieste, Italy}
\affiliation[d]{Department of Physics \& Wisconsin IceCube Particle Astrophysics Center, University of Wisconsin,
	Madison, WI 53706, U.S.A.}
\affiliation[e]{Dipartimento di Matematica e Fisica, 
Universit\`a di Roma Tre\\Via della Vasca Navale 84, 00146 Rome, Italy}	

\emailAdd{sanjib@iopb.res.in (ORCID:0000-0002-9714-8866)} 
\emailAdd{sudipta.d@iopb.res.in (ORCID:0000-0002-5508-7751)} 
\emailAdd{giarnetti.alessio@unioma3.it (ORCID:0000-0001-8487-8045)} 
\emailAdd{davide.meloni@uniroma3.it (ORCID:0000-0001-7680-6957)} 

\abstract
{Our knowledge on the active 3$\nu$ mixing angles 
($\theta_{12}$, $\theta_{13}$, and $\theta_{23}$) and the CP phase $\delta_{\mathrm{CP}}$ is becoming accurate
day-by-day enabling us to test the unitarity of the leptonic mixing 
matrix with utmost precision. Future high-precision long-baseline 
experiments are going to play an important role in this direction. 
In this work, we study the impact of possible non-unitary neutrino
mixing (NUNM) in the context of next-generation long-baseline experiments 
DUNE and T2HKK/JD+KD having one detector in Japan (T2HK/JD) and a second 
detector in Korea (KD). We estimate the sensitivities of these
setups to place direct, model-independent, and competitive 
constraints on various NUNM parameters. We demonstrate the possible 
correlations between the NUNM parameters,
$\theta_{23}$, and $\delta_{\mathrm{CP}}$.
Our numerical results obtained using only far detector data and supported by simple 
approximate analytical expressions of the oscillation probabilities 
in matter, reveal that JD+KD has better sensitivities for $|\alpha_{21}|$ 
and $\alpha_{22}$ as compared to DUNE, due to its larger statistics
in the appearance channel and less systematic uncertainties in the
disappearance channel, respectively. For $|\alpha_{31}|$, $|\alpha_{32}|$, and $\alpha_{33}$, 
DUNE gives better constraints as compared to JD+KD, due to its 
larger matter effect and wider neutrino energy spectrum.
For $\alpha_{11}$, both DUNE and JD+KD give similar bounds.
We also show how much the bounds on the NUNM parameters can be improved 
by combining the prospective data from DUNE and JD+KD setups. We find that due to zero-distance effects, the near detectors alone can also constrain $\alpha_{11}$, $|\alpha_{21}|$, and $\alpha_{22}$ in both these setups.
Finally, we observe
that the $\nu_\tau$ appearance sample in DUNE can improve the 
constraints on $|\alpha_{32}|$ and 
$\alpha_{33}$.
}


\arxivnumber{2111.00329}

\begin{document}
\maketitle
\flushbottom

\section{Introduction and motivation}
\label{sec:introduction}
The Standard Model (SM) of particle physics is the most successful 
gauge theory which has been rigorously tested in 
several laboratory experiments, including those which are being 
performed at the Large Hadron Collider at CERN in Geneva \cite{Evans:1129806}.
In spite of this huge success, the observed input parameters of the SM, 
such as the Higgs mass, the fermion masses and mixings, and 
the QCD theta term are quite offbeat and certainly require further
explanations~\cite{Dine:2000cj, Gonzalez-Garcia:2002bkq}. These pressing issues point towards the existence 
of a microscopic theory beyond the Standard Model which should address
the electroweak hierarchy problem~\cite{Jegerlehner:2013nna, tHooft:1979rat},
the flavor puzzle, and the strong CP problem \cite{McCullough:2018knz, Peccei:2006as,PhysRevLett.38.1440}. Various next-generation high-precision experiments
in the energy, intensity, 
and cosmic frontiers are expected to provide crucial information on these issues. 
In the intensity frontier, several high-precision
neutrino oscillation experiments are currently running and also in the pipeline to measure the 
mass-mixing parameters with unprecedented precision. Marvelous data from several pioneering neutrino experiments like Super-K~\cite{Super-Kamiokande:2010tar,Super-Kamiokande:2017yvm,Super-Kamiokande:2019gzr}, IceCube-DeepCore~\cite{IceCube:2017lak}, ANTARES~\cite{ANTARES:2018rtf}, Daya Bay~\cite{DayaBay:2018yms}, RENO~\cite{RENO:2018dro}, MINOS~\cite{MINOS:2013utc}, Tokai-to-Kamioka (T2K)~\cite{T2K:2019bcf}, and NuMI Off-axis $\nu_e$ Appearance (NO$\nu$A)~\cite{NOvA:2021nfi} have been decisive for our understanding of neutrino flavor mixing. Global fit analyses of the world neutrino data~\cite{NuFIT,Esteban:2020cvm,deSalas:2020pgw,Capozzi:2021fjo} have already been able to determine the values of the neutrino oscillation parameters with reasonable accuracy.
At present, the relative 1$\sigma$ precision on the mixing angles $\theta_{23}$, $\theta_{13}$, and $\theta_{12}$ lies in the range of ${\cal O}(3-7)\%$. 
For the mass-squared differences, the achieved relative 1$\sigma$ precision is around ${\cal O}(1-3)\%$.
Another important result that is being emerged from these global fit studies is that now we have an overall 1.6$\sigma$ hint for leptonic CP-violation  ($\sin\delta_{\mathrm{CP}}<0$)~\cite{Esteban:2020cvm,Capozzi:2021fjo}.

The upcoming medium-baseline reactor experiment Jiangmen Underground Neutrino Observatory (JUNO)~\cite{JUNO:2015zny,JUNO:2021vlw} is going to address the issue of neutrino mass hierarchy at high confidence level. JUNO will also improve our knowledge on the mixing angle $\theta_{12}$ and mass-squared differences $\Delta m^2_{21}$ and $\Delta m^2_{31}$.
The Fermilab-based Deep Underground Neutrino Experiment (DUNE)~\cite{DUNE:2020lwj,DUNE:2020jqi,DUNE:2021cuw,DUNE:2021mtg} is expected to resolve the issue of neutrino mass hierarchy and leptonic CP violation with a confidence level never achieved before. DUNE will measure the value of CP phase $\delta_{\mathrm{CP}}$ and atmospheric oscillation parameters $\theta_{23}$ and $\Delta m^2_{31}$ with high precision using its wide band neutrino beam, which provides information on oscillation parameters at several $L/E$ values.
Another major candidate in the race of next-generation high-precision long-baseline neutrino oscillation experiments is Tokai to Hyper-Kamiokande (T2HK) with one detector in Japan, which we refer as JD in the present work~\cite{Hyper-KamiokandeProto-:2015xww, Hyper-Kamiokande:2018ofw}. The possibility of having a second detector in Korea (KD) is also being explored actively. The combination of these two detectors (one in Japan and other in Korea) exposed to a common high-intensity, off-axis ($2.5^{\circ}$), narrow-band beam from J-PARC, is widely known as T2HKK~\cite{Hyper-Kamiokande:2016srs}, which we refer as JD+KD setup in the present study.
The JD setup with a relatively shorter baseline as compared to DUNE and high statistics is going to address the issue of leptonic CP violation with unmatched sensitivity and measure the value of $\delta_{\mathrm{CP}}$ quite precisely without facing the issue of fake CP violation due to matter effect. On the other hand, the KD setup having a baseline roughly around four times that of JD contains some information on Earth matter effect and measure $\delta_{\mathrm{CP}}$ around the second oscillation maximum with reasonable precision.

With the excellent foreseen precision on the neutrino mass-mixing parameters in the coming years, it will be interesting to explore whether tiny new physics effects beyond the standard three-flavor framework are present in the scenario and what are their impact on the neutrino oscillation probabilities~\cite{Fong:2016yyh}. In the present paper, in the context of high-precision long-baseline experiments, we study the possible consequences of the fact that the $3\times3$ active neutrino mixing matrix $N$ no longer respects the unitarity condition $N^\dagger N=I$. Theoretically, there are various neutrino mass models in the literature~\cite{Mohapatra:1986bd, Mohapatra:1979ia, Akhmedov:1995vm, Malinsky:2009gw}, which allows the possibility of non-unitary neutrino mixing (NUNM).
The most appealing ones are the so-called see-saw models, in which new heavy neutral leptons and/or scalars are introduced in the basic Standard Model to 
explain tiny neutrino masses.
 In these models, the standard 3$\times$3 active neutrino mixing matrix becomes a non-unitary sub-matrix of the larger mixing matrix (see discussion in Appendix of Ref.~\cite{Escrihuela:2015wra}). If the heavy lepton is very massive in the see-saw model~\cite{deGouvea:2015euy,Blennow:2016jkn, Escrihuela:2016ube}, then it can cause a departure from the unitarity of the order of 
 $10^{-3}$. The amount of non-unitarity can be even larger in the low-scale see-saw models~\cite{Akhmedov:1995ip,Malinsky:2009df}.
 
In this work, our aim is to study the sensitivity of various long-baseline experiments to probe the non-unitarity of the standard $3\times3$ active neutrino mixing matrix via oscillation, which allows us to constrain the NUNM parameters considering one at a time or all of them at the same time. Here, we adopt a completely model-independent approach to invoke this possible deviation from unitarity without relying on the underlying mechanism.
 
In the literature, many theoretical/phenomenological studies have been performed to investigate the possible signatures of non-unitarity of the three-flavor neutrino mixing matrix in the context of neutrino oscillation experiments. Considerable efforts have been made to study the possible impact of NUNM on the measurement of leptonic CP phase $\delta_{\mathrm{CP}}$, determination of the neutrino mass-hierarchy~\cite{Li:2015oal,Parke:2015goa,Ge:2016xya,Miranda:2016wdr,Dutta:2016czj,Pas:2016qbg}, and in
estimating the performance of current and future long-baseline experiments to constrain the various NUNM parameters~\cite{Antusch:2006vwa,Miranda:2019ynh,DeGouvea:2019kea,Escrihuela:2015wra,Fernandez-Martinez:2016lgt,Forero:2021azc,Ellis:2020hus,Coloma:2021uhq,Hu:2020oba,Blennow:2016jkn,Abe:2017jit,Dutta:2019hmb}.

In the present work, we study the potential of two next-generation long-baseline experiments DUNE and T2HKK (JD+KD), to constrain the NUNM parameters in a complete model-independent approach. We present the expected bounds, as obtained from these experiments individually as well as from their combination, assuming no priors in the fit procedure. 
Compared to other similar studies, we differ in several aspects; first of all, to carry on our numerical simulations, we use the most recent configuration of the DUNE~\cite{DUNE:2021cuw} and T2HKK~\cite{Hyper-Kamiokande:2016srs} experiments, also investigating the role of near detectors (ND) and $\tau$ neutrino detection in the case of DUNE. Then, for a better understanding of the obtained bounds as well as of the observed correlations between the standard oscillation parameters  $\theta_{23}$ and $\delta_{\mathrm{CP}}$ with the various NUNM parameters, we provide simple and useful analytical expressions of the relevant transition probabilities in the regime of small matter effects, which is a correct approximation for the neutrino facilities under investigation. 

 The manuscript is organized as follows. In Sec.~\ref{formulae}, we discuss the formalism and the parameterization of non-unitary neutrino mixing used in this work, and derive  simple approximate analytical expressions of the oscillation probability in $\nu_{\mu}\rightarrow\nu_{e}$ and $\nu_{\mu}\rightarrow\nu_{\mu}$ oscillation channels in the presence of standard matter interaction. In Sec.~\ref{sec:experiments}, we give a brief description of the future long-baseline experiments discussed in this work. Sec.~\ref{sec:sim-details} provides the details of the numerical simulations performed in our analysis and some results on the expected signal events. The main results of our study are presented in Sec.~\ref{results}, where we illustrate the correlation of the standard oscillation parameters $\theta_{23}$ and $\delta_{\mathrm{CP}}$ with various NUNM parameters and give the expected constraints achievable by DUNE and T2HKK separately as well as their combination. In Sec.~\ref{sec:ND}, we show the possible improvements in the bounds due to the presence of near detectors. 
 Section~\ref{sec:tau-analysis} shows the improved constraints on the NUNM parameters when we add $\nu_{\tau}$ events sample in the DUNE simulation. Our concluding remarks are discussed in Sec.~\ref{sec:conclusion}. Two appendices complete our work: in Appendix~\ref{Appendix:A}, we compare our analytical expressions of the oscillation probabilities and those computed numerically in the presence of non-unitarity. We also provide the analytical expressions of the  oscillation probabilities for all the remaining channels. In Appendix~\ref{sec:NU_marg}, we discuss in detail the impact on the bounds discussed in the main text of the simultaneous marginalization over all NUNM parameters.


\section{Oscillation probabilities in matter with non-unitary neutrino mixing}
\label{formulae}
The non-unitarity of the PMNS matrix can be parameterized in different ways~\cite{Antusch:2006vwa, Xing:2012kh,Flieger:2019eor,Bielas:2017lok,Ellis:2020ehi,Hu:2020oba}. 
One possibility, which turned out to be very useful in oscillation analysis, consists of factorizing the deviation from unitarity into a matrix $\alpha$ in such a way that the non-unitary neutrino mixing matrix $N$ is expressed as\footnote{We choose the convention in which the matrix $\alpha$, which invokes non-unitarity, is added to the identity matrix. Note that other phenomenological studies adopt the relation $N=(1-\alpha)\,U_{PMNS}$~\cite{Blennow:2016jkn}. Our results can be compared to the others just changing the sign in front of the diagonal elements.}:
\begin{eqnarray}\label{matriceN}
N=(I+\alpha)\,U_{PMNS}\,.
    \end{eqnarray}
    
In order to compare our numerical results with those already presented in the literature, we adhere here at the widely used lower triangular structure of the matrix $\alpha$, containing nine free parameters organized as follows:   
\begin{eqnarray}
\alpha=
\begin{pmatrix}
\alpha_{11} & 0 & 0 \\
|\alpha_{21}|e^{i \phi_{21}} & \alpha_{22} & 0 \\
|\alpha_{31}|e^{i \phi_{31}} & |\alpha_{32}|e^{i \phi_{32}} & \alpha_{33} 
\end{pmatrix}\, \, .
\label{eq:NU-triangle-mat}
\end{eqnarray}
This parameterization simplifies the oscillation probabilities and let the parameter $\alpha_{ij}$ to be the main source of non-unitarity for the oscillation channel $\nu_i\to\nu_j$ ($i,j = e,\mu,\tau$).

Bounds on the $\alpha_{ij}$ parameters have been recently computed, among others, in Ref.~\cite{Forero:2021azc} and reported for convenience in Table~\ref{alphabounds1}. These results have been obtained using data from the short-baseline experiments NOMAD and NuTeV, and the long-baseline experiments MINOS/MINOS+, T2K, and NO$\nu$A. 
For the off-diagonal NUNM parameters, the authors also used the triangular inequalities\footnote{Note that in this model, the diagonal parameters can only be negative.} $\alpha_{ij}\leq \sqrt{1-(1+\alpha_{ii})^2}\sqrt{1-(1+\alpha_{jj})^2}$~\cite{Escrihuela:2016ube}, which appear due to the assumption that the standard 3$\times$3 active neutrino mixing matrix is a non-unitary sub-matrix of a larger unitary mixing matrix.
  Note that in the present study, we do not take into account these inequalities in order to study the capability of long-baseline experiments alone to put bounds on these NUNM parameters in a model-independent fashion.

 \begin{table}
 	\centering
 	\begin{tabular}{|c|c|c|c|c|c|}
 		\hline\hline
 		$\alpha_{11}$ & $\alpha_{22}$ & $\alpha_{33}$ & $|\alpha_{21}|$ & $|\alpha_{31}|$ & $|\alpha_{32}|$\\ 
 		\hline
 		$<$ 0.031 &  $<$ 0.005 & $<$ 0.110 & $<$ 0.013 & $<$ 0.033 & $<$ 0.009  \\
 			\hline\hline
 	\end{tabular}
 	\mycaption{\label{alphabounds1}90\% confidence level (C.L.) limits on the NUNM parameters using data from various short-baseline and long-baseline  experiments, as obtained from the Ref.~\cite{Forero:2021azc}.} 
 \end{table}

 \begin{table}
 	\centering
 	\begin{tabular}{|c|c|c|c|c|c|}
 		\hline\hline
 		$\theta_{23}$&$\theta_{13}$&$\theta_{12}$&$\delta_{\mathrm{CP}}$&$\Delta m^2_{21} [\mathrm{eV^2}]$&$\Delta m^2_{31} [\mathrm{eV^2}]$ \\
 		\hline
 		$45^{\circ}$ & $8.61^{\circ}$&$33.6^{\circ}$& $-90^{\circ}$ &$ 7.39\times 10^{-5}$&$2.52\times 10^{-3}$ \\
 		\hline\hline
 	\end{tabular}
 	\mycaption{The benchmark values of the oscillation parameters used in our analysis. These values are consistent with the present best-fit
 		values as obtained in various global fit 
 		studies~\cite{NuFIT,Esteban:2020cvm,deSalas:2020pgw,Capozzi:2021fjo}.
 		We assume normal mass hierarchy (NH) of neutrino throughout this work.}
 	\label{table:vac}
 \end{table}
 
Taking into account the expression in Eq.~\ref{matriceN},  the complete effective neutrino propagation Hamiltonian in the mass-eigenstate is: 
\begin{eqnarray} \label{evol}
H= \frac{1}{2 E_\nu} \left[\left( \begin{array}{ccc}
                   0   & 0          & 0   \\
                   0   &  \Delta m^2_{21}  & 0  \\
                   0   & 0           &  \Delta m^2_{31}  
                   \end{array} \right)  +  N^{\dagger}
                  \left( \begin{array}{ccc}
            a_e + a_n      & 0 & 0 \\
            0  & a_n  & 0 \\
            0 & 0 & a_n
                   \end{array} 
                   \right) N\right]
\, .
\label{eq:matter}
\end{eqnarray}
As usual, the matter potential parameters are given by
$a_e= 2 \sqrt{2} E_\nu G_F N_e$ and $a_n= -\sqrt{2} E_\nu G_F N_n$ where, $N_e$ and $N_n$ are the electron and the neutron number densities, respectively. Note that in this framework, the neutral current (NC) matter potential is necessary since the non-unitarity of the matrix $N$ does not allow the subtraction of an identity matrix proportional to $a_n$. 
From the Schroedinger equation, the transition probability at a given baseline $L$ is obtained from the following expression:
\begin{eqnarray}
P_{\alpha\beta}=|(N e^{-i H L}N^\dagger)_{\beta\alpha}|^2.
\end{eqnarray}
The relevant probabilities for long-baseline experiments are the $\nu_{\mu}\rightarrow\nu_e$ appearance and $\nu_\mu\rightarrow\nu_\mu$ disappearance channels. We will discuss these probabilities here in details, while for the sake of completeness, we also quote the $\tau$ appearance and electron disappearance into the Appendix~\ref{Appendix:A}.

In order to get approximate analytical expressions for the transition probabilities,  we observe that the vacuum approximation cannot be sufficiently precise in  experiments like DUNE, since the matter effects can modify the appearance probability up to about 10\%. For this reason, we derive approximate analytical expressions in the presence of matter. We use perturbation theory in the small expansion
parameters ($r,\,s,$ and $a$) defined as follows:
\begin{equation}
\label{expansion}
\sin \theta_{13} = \frac{r}{\sqrt{2}}\,, \qquad \sin \theta_{12} = \frac{1}{\sqrt{3}}(1+s)\,,\qquad \sin \theta_{23} =  \frac{1}{\sqrt{2}}(1+a)\,,
\end{equation}
where, $r,\, s$, and $a$ represent the deviation from the tri-bimaximal mixing values of the neutrino mixing parameters, namely, $\sin \theta_{13}=0$, $\sin \theta_{23} = 1/\sqrt{2}$, and $\sin \theta_{12} = 1/\sqrt{3}$~\cite{King:2007pr,Pakvasa:2007zj}. Given the recent global fit of neutrino oscillation data, we can assume that $r,s,a \sim {\cal O}(0.1)$ and we can further expand them up to the second order~\cite{NuFIT,Esteban:2020cvm,deSalas:2020pgw,Capozzi:2021fjo}. 
To further simplify the notation, we also introduce  $\Delta_{31} = \Delta m_{31}^2 L/ 4E_\nu$,  $\Delta_e = a_e L/ 4E_\nu$ and $\Delta_n = a_n L/ 4E_\nu$; at the DUNE peak energy, namely, $E_\nu = 2.5$ GeV, $\Delta_e \sim 0.36$ and $\Delta_n \sim 0.18$, we can further expand in the small matter potentials up to the first order. Note that for the other experimental facilities discussed in this paper, this approximation is even better; in fact, for the Tokai to Hyper-Kamiokande (T2HK)  setup with a far detector in Japan (JD),  at beam energy of $E_\nu =0.6$ GeV, we have $\Delta_e \sim 0.08$ and $\Delta_n \sim 0.04$, while with a second  in Korea (KD) at a distance of 1100 km from the source with same energy, we get $\Delta_e \sim 0.30$ and $\Delta_n \sim 0.15$. Also, we use one mass scale dominance (OMSD) approximation ($\Delta_{31}\,>>\,\Delta_{21}$, where $\Delta_{21}=\Delta m^2_{21}L/4E_{\nu}$) in our derivation; which is a valid approximation in the atmospheric regime.

In the case of the $\nu_{\mu}\rightarrow\nu_{e}$ appearance probability, we thus obtain:
\begin{eqnarray}
P_{\mu e} &=& \left(\frac{r^2}{\Delta_{31}}\right)\, \sin \Delta_{31} \left[(\Delta_{31}+2 \Delta_{e}) \sin \Delta_{31}-2 \Delta_{31} \Delta_{e} \cos\Delta_{31}\right]\nonumber + \\
   && \left(\frac{2 |\alpha_{31}|  r}{\Delta_{31}}\right) \Delta_n \sin\Delta_{31} \left[\cos (\delta_{\mathrm{CP}}-\phi_{31}) \sin\Delta_{31}-\Delta_{31}
   \cos (\delta_{\mathrm{CP}}+\Delta_{31}-\phi_{31})\right] \nonumber + \\
&& \left(\frac{|\alpha_{21}| r}{\Delta_{31}} \right)  \,\left\{
\sin\Delta_{31}\left[2 \Delta_{31}   (\Delta_e+\Delta_n) \cos (\delta_{\mathrm{CP}}+\Delta_{31}  -\phi_{21})-\Delta_n \sin (\delta_{\mathrm{CP}}-\Delta_{31}  -\phi_{21})\right]+\right. \nonumber 
\\
&& \left.\sin (\delta_{\mathrm{CP}}+\Delta_{31}  -\phi_{21}) \left[(-2 \Delta_{31} -2 \Delta_e+\Delta_n) \sin\Delta_{31}+2 \Delta_{31}  \Delta_e  \cos\Delta_{31}\right]\right\}  + \nonumber\\
&& \left(\frac{|\alpha_{21}||\alpha_{31}|}{\Delta_{31} }\right) \Delta_n
\left[-2 \Delta_{31}  \sin (\phi_{21}-\phi_{31})+\cos (2 \Delta_{31} -\phi_{21}+\phi_{31})-\cos (\phi_{21}-\phi_{31})\right] \nonumber + \\ &&
   \left(\frac{|\alpha_{21}|^2}{\Delta_{31}}\right)\left[\Delta_{31}-\Delta_n (1-\cos 2  \Delta_{31})\right]\label{eq:pme_anylit_main}\,.
\end{eqnarray}
From the above expression, it is clear that the $\nu_{\mu}\rightarrow\nu_{e}$ appearance probability strongly depends on $|\alpha_{21}|$ and $|\alpha_{31}|$. 
 The parameter $|\ame|$ survives in the vacuum case while $|\alpha_{31}|$ always appears with the matter potential $\Delta_n$. This essentially means that an experiment in which the matter effect is not negligible is able to put strong bounds also to $|\alpha_{31}|$ which, otherwise, would not be accessible by $P_{\mu e}$. 
Note that due to the loss of unitarity property of the neutrino mixing, some terms remain non-zero in $\nu_{\mu}\rightarrow\nu_{e}$ appearance probability expression even when the neutrino propagation length $L$ is zero, which is known as zero-distance effect: 
\begin{eqnarray}
P_{\mu e}^{L = 0} &\sim& |\alpha_{21}|^2 \label{zerodistancemue}\,.
\end{eqnarray}
So, it is clear that even the near detectors (ND) of long-baseline experiments could contribute to the bounds of non-unitarity parameters, as it will be discussed in Sec.~\ref{sec:ND}. Finally, we point out that the vacuum limit of 
Eq.~\ref{eq:pme_anylit_main} assumes a particularly simple expression: 
\begin{eqnarray}
P_{\mu e}^{vac}&=&
|\alpha_{21}|^2+r^2 \sin ^2\Delta_{31}-2 |\alpha_{21}| r 
\sin\Delta_{31} \sin (\delta_{\mathrm{CP}}+\Delta_{31}-\phi_{21})
\label{pmuevacuum} \,,
\end{eqnarray}
which, in the limit of vanishing solar mass difference, agrees with the results presented in Ref.~\cite{Escrihuela:2015wra}.

\begin{figure}[h!]
\centering
\includegraphics[width=1.05\textwidth]{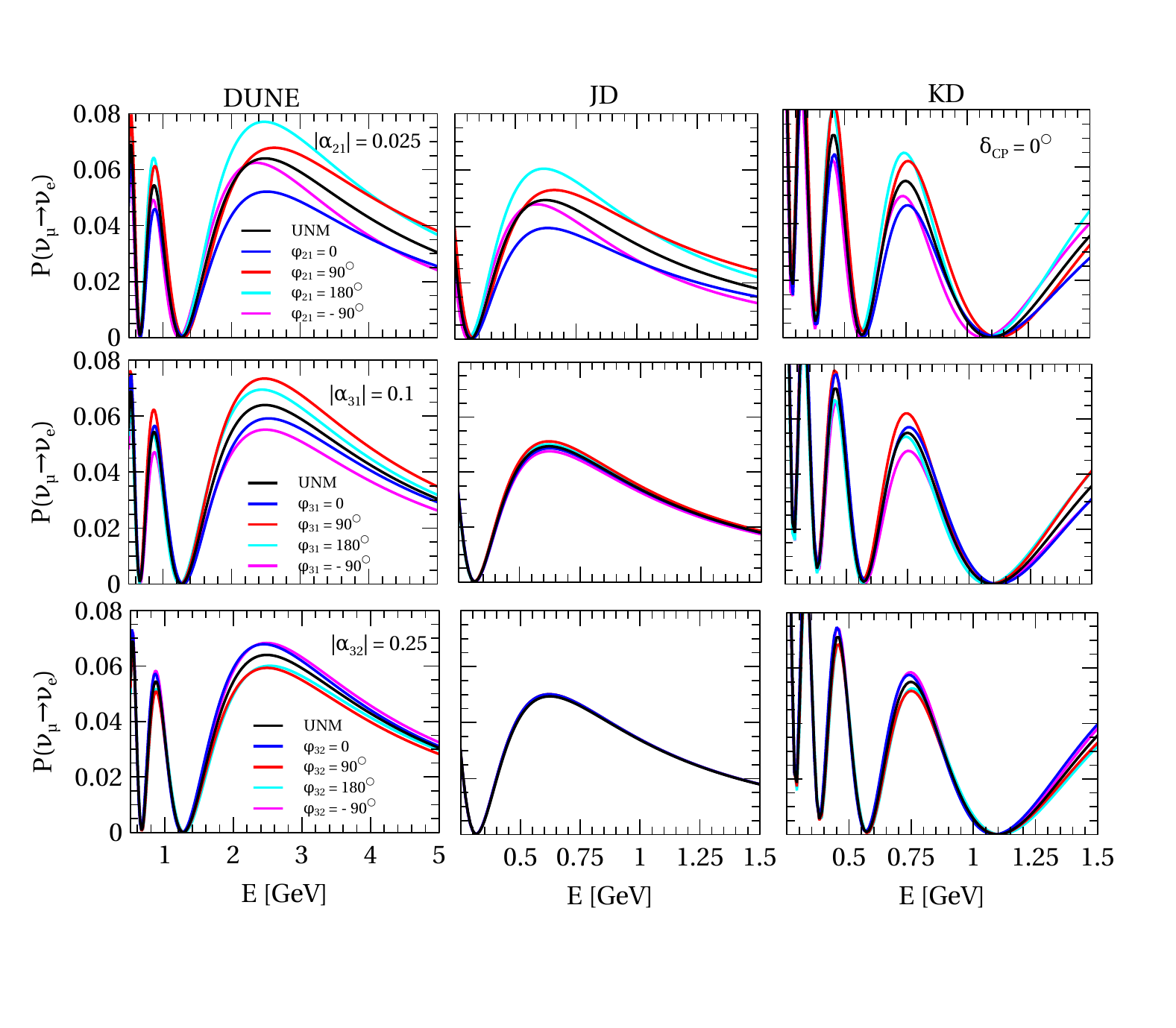}
\vspace{-0.8cm}
\mycaption{$\nu_{\mu}\rightarrow\nu_{e}$ appearance probability as a function of energy in the presence of off-diagonal NUNM parameters. Left, middle, and right columns correspond to the baselines of 1300 km (DUNE), 295 km (JD), and 1100 km (KD), respectively. The four colored curves correspond to four benchmark values of the phases associated with off-diagonal NUNM parameters: $0^{\circ}$, $90^{\circ}$, $180^{\circ}$, and $-90^{\circ}$. We consider $\delta_{\mathrm{CP}} = 0^{\circ}$ and $\sin^2\theta_{23}=0.5$. The values of the other oscillation parameters are taken from Table~\ref{table:vac}. }
\label{fig:prob_mue}
\end{figure}
\begin{figure}[h!]
\centering
\includegraphics[width=\textwidth]{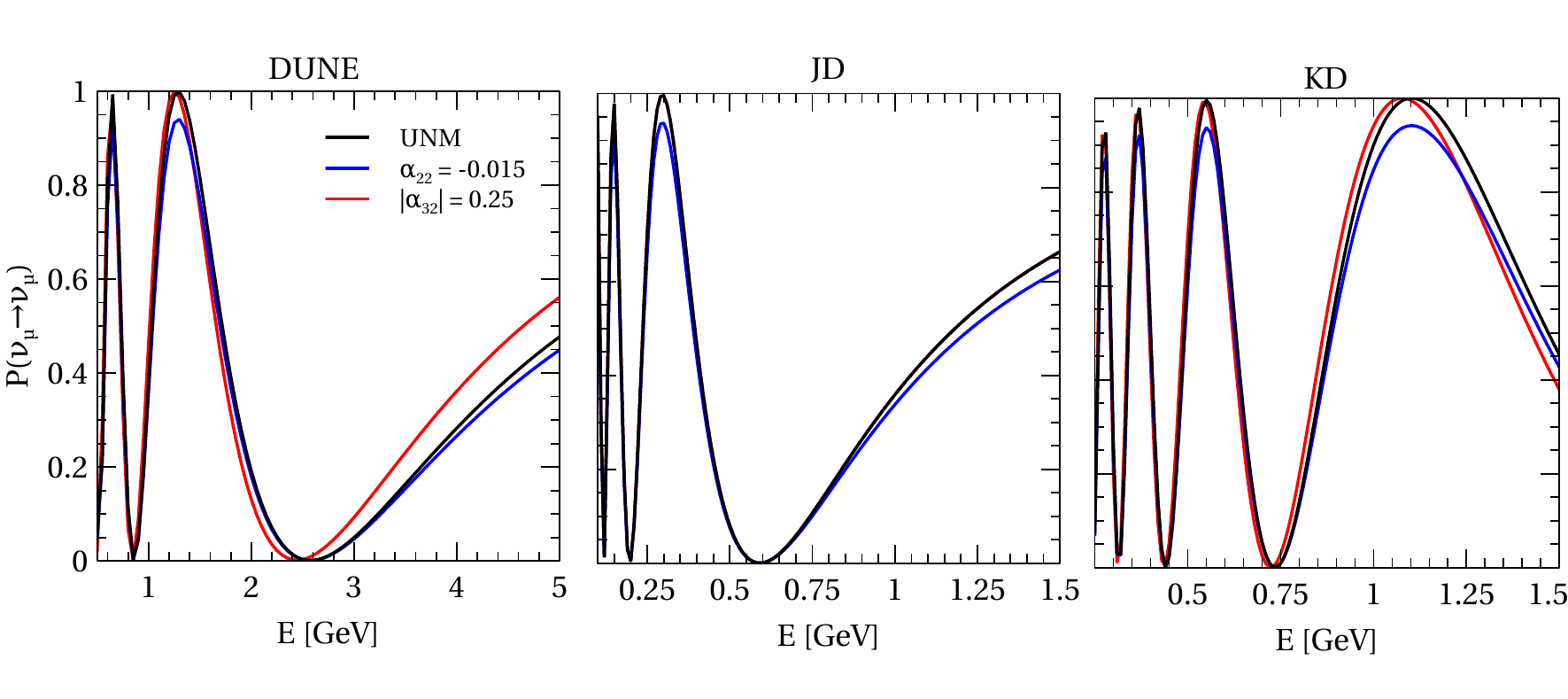}
\vspace{-0.5cm}
\mycaption{$\nu_{\mu}\rightarrow\nu_{\mu}$ disappearance probabilities as a function of neutrino energy in the presence of the NUNM parameters $\alpha_{22}$ and $|\alpha_{32}|$ assuming $\phi_{32}=0$ one at a time. Left, middle, and right columns correspond to the baselines of 1300 km (DUNE), 295 km (JD), and 1100 km (KD), respectively. We consider $\delta_{\mathrm{CP}} = 0^{\circ}$ and $\sin^2\theta_{23}=0.5$. The values of the other oscillation parameters are taken from Table~\ref{table:vac}.}
\label{fig:prob_mumu}
\end{figure}

In order to assess the modifications in the $\nu_{\mu}\rightarrow\nu_{e}$ appearance probabilities caused by the presence of non-unitary neutrino mixing,
in Fig.~\ref{fig:prob_mue}, we show the exact  $\nu_{\mu}\rightarrow\nu_{e}$ appearance probabilities as a function of energy, obtained numerically using the General long-baseline Experiment Simulator (GLoBES) software~\cite{Huber:2004ka,Huber:2007ji}. In the extreme left column, we consider a baseline of 1300 km for DUNE. In the middle column, we show the results for the JD baseline of 295 km.
 In the extreme right column, we deal with the KD setup having a baseline of 1100 km. Note that for both JD and KD, we consider the neutrino energy range of 0 to 1.5 GeV having a peak around 0.6 GeV. In every rows, we switch-on one off-diagonal NUNM parameters  at a time, while maintaining  others to zero. The effect of $|\ame|$ is shown in the top panels, $|\ate|$ in the middle panels, and $|\atm|$ in the bottom panels. In each panel, the solid black curves correspond to the probabilities in the unitary neutrino mixing (UNM) case, while the colored curves correspond to the NUNM cases with four benchmark values of the phases associated with each off-diagonal NUNM parameters, as reported in the legend.
As we can see,  the impact of the NUNM parameter $|\ame|$ (top panels) is comparatively larger than the other two off-diagonal NUNM parameters $|\ate|$ and $|\atm|$ even though the strength of the $|\ame|$ is much smaller than the other two. This feature is more clear for JD in the middle columns. In fact,  from the approximated analytical expression of $\nu_{\mu}\rightarrow\nu_{e}$ appearance probability in Eqs.~\ref{eq:pme_anylit_main} and \ref{pmuevacuum}, we see that  only terms containing $|\ame|$ survive in vacuum, while the effect of $|\ate|$ and $|\atm|$ is linked to $\Delta_e$ and $\Delta_n$ which are very small at the considered baseline. This is not the case for DUNE ($L \simeq 1300$ km) where, given the largest baseline under consideration, $\Delta_e$ and $\Delta_n$ are no longer negligible and the impact of $|\ate|$ and $|\atm|$ on $P_{\mu e}$ is of the same order as $|\ame|$.

For the $\nu_\mu\rightarrow\nu_{\mu}$ disappearance channel, we get:
\begin{eqnarray}
 P_{\mu \mu} &=&\cos^2\Delta_{31} (1+ 4 \alpha_{22})-2 |\alpha_{32}| \Delta_n \sin2\Delta_{31} \cos\phi_{32} + 4 a^2 \sin^2\Delta_{31} + \nonumber \\
 && 2 |\alpha_{21}|^2 \cos\Delta_{31} \left[2 \sin\Delta_{31}(\Delta_e+\Delta_n)+\cos\Delta_{31}\right] + 6 \alpha_{22}^2 \cos^2\Delta_{31} - \nonumber \\
 &&2 |\alpha_{21}| r \sin (2 \Delta_{31}) (\Delta_e+\Delta_n) \cos(\delta_{\mathrm{CP}}-\phi_{21}) + \nonumber\\
 &&\left(\frac{8 a}{\Delta_{31}}\right)(\alpha_{22}-\alpha_{33}) \Delta_n \sin \Delta_{31} (\sin \Delta_{31}-\Delta_{31} 
 \cos \Delta_{31}) \nonumber \\
 &&-2 |\alpha_{21}| |\alpha_{31}| \Delta_n  \sin2 \Delta_{31} \cos (\phi_{21}-\phi_{31})
 -8 \alpha_{22} |\alpha_{32}| \Delta_n\sin2\Delta_{31} \cos\phi_{32} \nonumber \\
 &&-2 |\alpha_{32}| \alpha_{33} \Delta_{n} \sin2\Delta_{31}\cos\phi_{32}\,, \label{eq:pmm_anylit_main}
\end{eqnarray}
from which we learn that all the NUNM parameters $\alpha_{ij}$ but $\alpha_{11}$ enter into the probability expression. Also, in this case, we get the zero-distance expression of the $\nu_{\mu}\rightarrow\nu_{\mu}$ survival probability, given by:
\begin{eqnarray} \label{pmmzero}
P_{\mu\mu}^{L = 0} &\sim& 1+ 2 |\alpha_{21}|^2+6 \alpha_{22}^2+4 \alpha_{22}\,,
\end{eqnarray}
and the vacuum approximation (also in agreement with Ref.~\cite{Escrihuela:2015wra} in the limit of vanishing $\Delta m^2_{21}$):
\begin{eqnarray}
P_{\mu\mu}^{vac}&=& \cos^2\Delta_{31} \left(1+2 |\alpha _{21}|^2 +4 \alpha_{22}+ 6\alpha_{22}^2 \right) + 4 a^2 \sin ^2\Delta_{31}\,. \label{pmmvacuum}
\end{eqnarray}
In Fig.~\ref{fig:prob_mumu}, we show the exact $\nu_{\mu}\rightarrow\nu_{\mu}$ oscillation probabilities as a function of energy for the baseline lengths corresponding to DUNE (left panel), JD (middle panel), and KD (right panel) setups. In each panel, the black solid curves  correspond to the UNM case, while the red and blue curves show the presence of $\amm$ and $\atm$, respectively with strength reported in the legend\footnote{Even though the analytical expression of $P_{\mu\mu}$ reported in Eq.~\ref{eq:pmm_anylit_main} shows the presence of other NUNM parameters,
we have numerically checked that they do not have any significant impact.}, one at a time. The impact of these two NUNM parameters can be understood from our approximated analytical expressions in Eq.~\ref{eq:pmm_anylit_main}.
When matter effects are negligible (for example, in the middle panel of Fig.~\ref{fig:prob_mumu}), we expect that the parameter $\amm$ 
dominates the deviation from UNM since it appears already at first order  in 
$P_{\mu\mu}$. This remains true when matter parameters are switched-on; the relevant difference compared to the vacuum case relies on the fact that also $|\atm|$ enter at first order, although suppressed by $\Delta_{n}$. Thus, we expect that  
for DUNE and KD, one can see deviation from the UNM predictions,  as visible in Fig.~\ref{fig:prob_mumu}. Note that the impact of $\alpha_{32}$ is amplified by the larger benchmark value compared to the choice for $\alpha_{22}$.

\section{Key features of the experiments}
\label{sec:experiments}

Long-baseline experiments provide a clear environment to study the phenomenology of neutrino oscillations. Indeed, the possibility to fix the baseline and have a focused neutrino beam allow us to have different features of the experiments under control. The next-generation long-baseline experiments will increase the neutrino events statistics with the help of high precision detectors. Two experiments currently under construction are DUNE and T2HK. For the latter is being taken into account the possibility of a second far detector placed approximately at the second oscillation maximum. These two complementary experiments, which will be described in more details in the following subsections, are going to improve the measurements of the oscillation parameters and the searches of new physics effects in a significant way. In the non-unitarity framework, the unprecedented statistics collected by the detectors of the two experiments, as well as the possibility to look for $\nu_\mu$ disappearance, $\nu_e$ appearance but also $\nu_\tau$ appearance in DUNE, will provide the chance to bound at a very good level all the non-standard parameters.

\subsection{DUNE}

DUNE (Deep Underground Neutrino Experiment) is a next-generation long-baseline experiment which will play an important role in solving the existing puzzles in neutrino oscillation as well as in increasing the precision on the measured neutrino oscillation parameters. It will use an on-axis, high-intensity, wide-band neutrino beam traveling a distance of 1284 km from Fermilab to South Dakota. For this baseline, they consider an average matter density of $2.85$ g/cm$^{3}$.
 The far detector (FD) is a liquid argon time projection chamber (LArTPC) of 40 kt fiducial mass placed underground at the Homestake mine. 
 According to the recent Technical Design Report (TDR)~\cite{DUNE:2021cuw}, DUNE will use a proton beam having 1.2 MW power which will deliver $1.1\times10^{21}$ proton on target (P.O.T.) per year. DUNE considers a total run-time of 10 years with 5 years each in neutrino and antineutrino modes. However,
 following most of the existing studies related to DUNE, we present our results using a total run-time of 7 years with 3.5 years each in neutrino and antineutrino modes.
Beside the far detector, the possibility to have three modules of near detectors (ND) has been proposed~\cite{DUNE:2021tad}. The first one would be a liquid Argon Time Projection chamber (TPC) situated at a distance of 574 m from the source. It will measure the flux and cross-section of the neutrinos. The second one would be a multi purpose detector (MPD) equipped with a magnetic spectrometer with one ton High-Pressure Gaseous Argon TPC, which will be useful to study possible new physics signals. The third one would be 
the System for On-Axis Neutrino Detection (SAND). It will be made up of a former KLOE magnet and a calorimeter which will track the outgoing particles.

\subsection{T2HK (JD) and T2HKK (JD+KD)}

T2HK (Tokai to Hyper-Kamiokande) is another promising next-generation long-baseline experiment which will play a very important role in $\delta_{\mathrm{CP}}$ measurement as well as in the study of various BSM physics~\cite{Kelly:2017kch,Agarwalla:2018nlx,Choubey:2017ppj}. In the T2HK setup~\cite{Abe:2011ts,Hyper-KamiokandeWorkingGroup:2014czz,Hyper-KamiokandeProto-:2015xww}, an intense neutrino beam from the J-PARC proton synchroton facility will be detected at the Hyper-Kamiokande (HK) detector situated at a distance of 295 km from the source. The power of the proton beam at the source is 
1.3 MW which will produce $27\times10^{21}$ P.O.T. in its total 10 years run-time.  The detector is a water Cherenkov (WC) detector with 187 kt of fiducial mass.
To have an equal contribution from the neutrino and the antineutrino signal events, the proposed running time of the experiment is 2.5 years (25\% exposure) in neutrino mode and 7.5 years (75\% exposure) in antineutrino mode.
The HK detector will be placed at $2.5^{\circ}$ off-axis angle from the neutrino beam-line to receive a narrow band beam with energy peaked at around the first oscillation maximum ($\sim 0.6$ GeV).

There is also a proposal~\cite{Seo:2019dpr} to have another identical detector which will be situated in Korea, 1100 km far from J-PARC. We assume that this detector will also be placed at an off-axis angle of $2.5^{\circ}$ from the neutrino beamline (similar to JD) and operate around the second oscillation maximum with a baseline of 1100 km and an average neutrino energy of 0.6 GeV. Combination of the T2HK setup along with the detector in Korea is known as T2HKK. However, in our work, we call the Hyper-Kamiokande as the Japanese detector (JD) and the detector in Korea as the Korean detector (KD). For both JD and KD baselines, we assume an average matter density of 2.8 g/cm$^3$.
 In the following, we will show our numerical results on the sensitivity to the various NUNM parameters separately for each detectors as well as for their combination.

In addition to these two detectors, two more near detectors, namely, the ND280 detector and Intermediate Water Cherenkov detector (IWCD), have been proposed. ND280 will be located at 280 m from the source with the same off-axis angle as the far detectors. It will be helpful to measure the flux of the unoscillated neutrino beam and study the neutrino cross-sections~\cite{T2K:2019bbb}. IWCD is a water Cherenkov detector of mass of 1 kt and possibly located at a distance of 1 km from the source~\cite{Drakopoulou:2017qdu, Wilson:2020trq}. One advantage of this detector is that it can be moved vertically to take data at different off-axis angles.  
In Table~\ref{tab:exp_details}, we summarize the relevant information of all the experimental setups discussed in this section.

\begin{table}
\centering
\begin{tabular}{|c|c|c|}
\hline \hline
 & DUNE & JD/KD\\ 
\hline
Detector Mass& 40 kt LArTPC & 187 kt WC (each)  \\
\hline
Baseline & 1300 km & 295/1100 km \\
\hline
Proton Energy & 120 GeV & 80 GeV \\
\hline
Beam type & Wide-band, on-axis & Narrow-band, off-axis ($2.5^{\circ}$)\\
\hline
Beam power & 1.2 MW & 1.3 MW \\
\hline
P.O.T./year& $1.1\times10^{21}$ & $2.7\times10^{21}$\\ 
\hline
Run time ($\nu+\bar{\nu}$) & 3.5 yrs + 3.5 yrs & 2.5 yrs + 7.5 yrs  \\
\hline\hline
\end{tabular}
\mycaption{Essential features of DUNE~\cite{DUNE:2021cuw} and JD/KD~\cite{Hyper-Kamiokande:2016srs}  experiments used in our simulation.}
 \label{tab:exp_details}
\end{table}

\section{Simulation details and discussion at the event level}
\label{sec:sim-details}

In order to perform our numerical simulations, we use the General long-baseline Experiment Simulator (GLoBES) package~\cite{Huber:2004ka,Huber:2007ji} along with the plug-in MonteCUBES~\cite{Blennow:2016jkn}. For the standard oscillation parameters, we use the value as given in Table~\ref{table:vac}. For the simulation of the DUNE experiment, we use the  most recent GLoBES configuration files as given in Ref.~\cite{DUNE:2021cuw}. The simulation details of T2HK and T2HKK setups are taken from Ref.~\cite{Hyper-Kamiokande:2016srs}. In Table~\ref{tab:exp_details}, we summarize the relevant information of the two experiments considered in our simulation.
Note that while showing the sensitivity of the FDs in constraining various NUNM parameters, we do not take into account the NDs in our simulation. The only exception is in Sec.~\ref{sec:ND}, where we discuss the possible improvement in the bounds due to presence of near detectors without considering any possible correlations among FDs and NDs (see Sec.~\ref{sec:ND} for further details). 
Both DUNE and T2HKK are expected to have access to two oscillation channels, namely the $\nu_\mu$ ($\bar{\nu}_\mu$) disappearance and the $\nu_e$ ($\bar{\nu}_e$) appearance channels. In particular:
\begin{itemize}
    \item for DUNE, backgrounds to the appearance channel are the $\nu_e$ beam contamination and the misidentified $\nu_\mu$, $\nu_\tau$, and NC events. The signal systematic normalization error have been chosen to be 2\%. For the disappearance channel, background to the signal are misidentified $\nu_\tau$ and NC events. The signal error is 5\%. Efficiency functions as well as smearing matrices have been provided by the DUNE collaboration in Ref.~\cite{DUNE:2021cuw};
    \item for the two T2HKK far detectors JD and KD, background in the appearance channels are the $\nu_e$ from the beam contamination and misidentified $\nu_\mu$ and NC events. Signal systematic uncertainties are 5\% normalization and 5\% calibration errors. In the disappearance channel, backgrounds to the signal are misidentified $\nu_e$ and NC events. Systematic uncertainties are 3.5\% normalization and 5\% calibration errors. Efficiencies and energy resolutions are taken from the Ref.~\cite{Hyper-Kamiokande:2016srs}.
\end{itemize}

\begin{table}

	\centering
	
	\begin{tabular}{|ccc|*{6}{c|}}
		
		\hline\hline
		
		\multicolumn{3}{|c|}{\multirow{2}{*}{}} & \multicolumn{3}{|c}{$\nu_e$ appearance} & \multicolumn{3}{|c|}{$\bar\nu_e$ appearance}\\ 
		
		\cline{4-9}
		
		\multicolumn{3}{|c|}{} & DUNE & JD & KD & DUNE & JD & KD \\
		
		\hline\hline
		
		\multicolumn{3}{|c|}{UNM} & 1259 & 1836 & 169 & 221 & 767 & 31 \\
		
		\cline{1-9}
		
		\hline\hline
		
		\multicolumn{3}{|c|}{NUNM} & DUNE & JD & KD & DUNE & JD & KD \\
		\hline\hline
		
		\multicolumn{3}{|c|}{$|\alpha_{21}|$ (= 0.025)} & 1328 & 1893 & 169 & 232 & 756 & 29 \\
		\hline
		\multicolumn{3}{|c|}{$|\alpha_{31}|$ (= 0.1)} & 1420 & 1893 & 187  & 264 & 754 & 30\\
		\hline
		\multicolumn{3}{|c|}{$|\alpha_{32}|$ (= 0.25)} & 1300 & 1855 & 172 & 213 & 756 & 30\\
		\hline
		\multicolumn{3}{|c|}{$\alpha_{11}$ (= -0.02)} & 1203 & 1761 & 162 & 214 & 738 & 29 \\
		\hline
		\multicolumn{3}{|c|}{$\alpha_{22}$  (= -0.015)} & 1223 & 1782 & 164 & 215 & 744 & 30 \\
		\hline
		\multicolumn{3}{|c|}{$\alpha_{33}$ (= -0.15)} & 1208 & 1817 & 168  & 227 & 779 & 30\\
		
		\hline
		\hline
		
	\end{tabular}
	
	\mycaption{Comparison of the total signal rate for the $\nu_{e}$ and $\bar{\nu_{e}}$ appearance channels in DUNE, JD, and KD setups in UNM case as well as in presence of various NUNM parameters. The relevant features of these facilities are given in Table~\ref{tab:exp_details}. The values of the standard oscillation parameters used to calculate event rate are quoted in Table~\ref{table:vac}. The phases associated with the off-diagonal NUNM parameters are considered
		to be zero.}
	
	\label{tab:total_events_app}
	
\end{table}
The number of expected signal events for both channels simulated here are summarized in Tables~\ref{tab:total_events_app} and \ref{tab:total_events_dis}, where the cases of UNM and NUNM (for some benchmark values of the $\alpha_{ij}$ parameters) are reported.
The impact of NUNM parameters on the number of events is fully in agreement with our analytical discussions. First of all, as shown in Eq.~\ref{eq:pme_anylit_main}, the appearance channel is mainly influenced by $|\alpha_{21}|$, even in vacuum. This reflects in an  enhancement of the number of events by roughly 5\% in both JD and DUNE. 
On the other hand, the NUNM parameters $|\alpha_{31}|$ and $\alpha_{33}$ 
are also relevant but they are coupled to the matter potentials, so we expect them to be relevant primarily for DUNE, where matter effects are more important: in fact, $|\alpha_{31}|$  causes an increase in the number of events up to 10\%, while $\alpha_{33}$ provokes a small but visible reduction of the order of  4\%.
Finally, some impact on the number of signal events is also given by $\alpha_{11}$, even though it only appears at higher orders in our perturbative expansion and has not been displayed (but it present in the vacuum probabilities reported in Ref.~\cite{Escrihuela:2015wra}). 
Note that the number of $\nu_e$ and $\bar{\nu}_e$ events in KD is only slightly influenced by the NUNM parameters due to fact that the experiment works close to the second oscillation maximum of the atmospheric oscillation ($\nu_{\mu}\rightarrow\nu_{\tau}$), where the $\nu_{\mu}\rightarrow\nu_e$ appearance probability approached one of its minima and the effects of new physics are suppressed.

For the disappearance channel, the parameter $\alpha_{22}$, which enters at the first perturbative order in Eq.~\ref{eq:pmm_anylit_main}, produces  a reduction of about 6\% in the number of events for all three detectors. This can be roughly understood from the fact that the standard disappearance  probability is multiplied by $4\alpha_{22}=0.06$, which causes a reduction by a similar factor in the number of events. The other relevant NUNM parameter is $|\alpha_{32}|$ which, being coupled to matter potential in Eq.~\ref{eq:pmm_anylit_main}, can cause a $\sim$ 6\% increase of events especially in DUNE. The other parameters at their benchmark values only have a negligible impact on the number of disappearance events.

\begin{table}
	
	\centering
	
	\begin{tabular}{|ccc|*{6}{c|}}
		
		\hline\hline
		
		\multicolumn{3}{|c|}{\multirow{2}{*}{}} & \multicolumn{3}{|c}{$\nu_\mu$ disappearance} & \multicolumn{3}{|c|}{$\bar\nu_\mu$ disappearance}\\ 
		
		\cline{4-9}
		
		\multicolumn{3}{|c|}{} & DUNE & JD & KD & DUNE & JD & KD \\
		
		\hline\hline
		
		\multicolumn{3}{|c|}{UNM} & 10359 & 9064 & 1266 & 6034 & 8625 & 1144 \\
		
		\cline{1-9}
		
		\hline\hline
		
		\multicolumn{3}{|c|}{NUNM} & DUNE & JD & KD & DUNE & JD & KD \\
		\hline\hline
		
		\multicolumn{3}{|c|}{$|\alpha_{21}|$ (= 0.025)} & 10371 &9074 & 1264 & 6045 & 8640 & 1149 \\
		\hline
		\multicolumn{3}{|c|}{$|\alpha_{31}|$ (= 0.1)} & 10351 & 9062 & 1261  & 6035 & 8627  & 1168\\
		\hline
		\multicolumn{3}{|c|}{$|\alpha_{32}|$ (= 0.25)} & 10978 & 9203 & 1255 & 6005 & 8467 & 1158\\
		\hline
		\multicolumn{3}{|c|}{$\alpha_{11}$ (= -0.02)} & 10359 & 9064 & 1266 & 6034 & 8625 & 1145 \\
		\hline
		\multicolumn{3}{|c|}{$\alpha_{22}$  (= -0.015)} & 9748 & 8531 & 1192 & 5681 & 8120 & 1077 \\
		\hline
		\multicolumn{3}{|c|}{$\alpha_{33}$ (= -0.15)} & 10406 & 9077 &  1268 & 6040 & 8619 & 1145 \\
		\hline
		\hline
		
	\end{tabular}
	
	\mycaption{Comparison of the total signal rate for the $\nu_{\mu}$ and $\bar{\nu_{\mu}}$ disappearance channels in DUNE, JD, and KD setups in UNM case as well as in presence of various NUNM parameters. The relevant features of these facilities are given in Table~\ref{tab:exp_details}. The values of the standard oscillation parameters used to calculate event rate are quoted in Table~\ref{table:vac}. The phases associated with the off-diagonal NUNM parameters are considered
			to be zero.}
	
	\label{tab:total_events_dis}
	
\end{table}

\begin{figure}[h!]
\centering
\includegraphics[width=\textwidth]{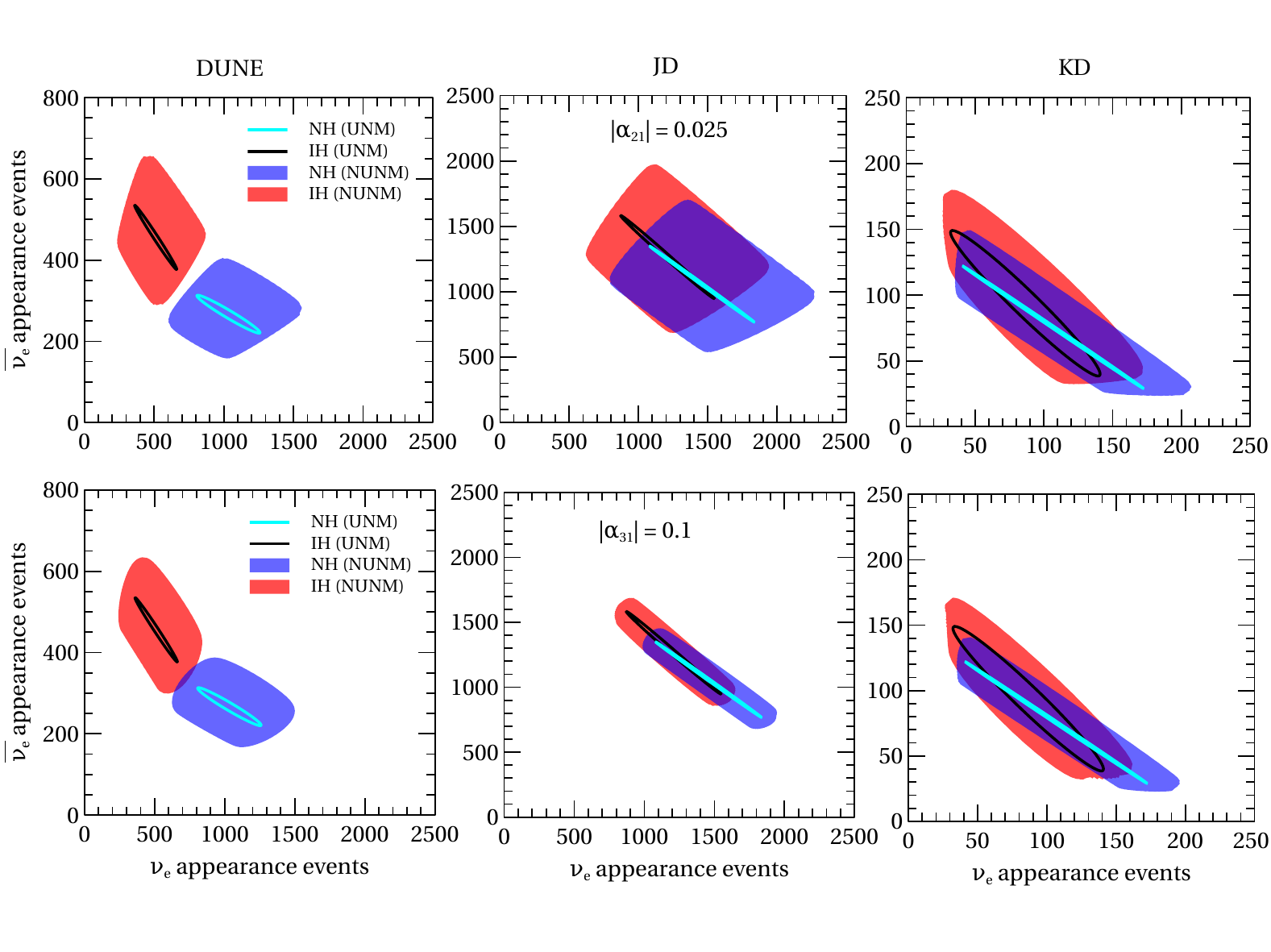}
\mycaption{Bi-event plots for the DUNE (left panel), JD (middle panel), and KD (right panel). The ellipses in each panel are obtained by varying the CP phase $\delta_{\mathrm{CP}}$ in the range $[-180^{\circ},\, 180^{\circ}]$ assuming $\alpha_{ij} = 0$. The colored blobs in the upper  panels show the bi-events in the presence of $|\alpha_{21}|$ with a strength of 0.025, while the lower panels depict the impacts of $|\alpha_{31}|$ with a strength of 0.1. To obtain the blobs with non-zero $\alpha_{ij}$,
	we vary the associated NUNM phases and the CP phase $\delta_{\mathrm{CP}}$ in their allowed range of $-\,180^{\circ}$ to $180^{\circ}$.
	 The values of the other standard three-flavor oscillation parameters are taken from Table~\ref{table:vac}.}
\label{fig:bi-event}
\end{figure}

In Fig.~\ref{fig:bi-event}, we show the relation between the appearance events in neutrino and antineutrino modes at the three detectors discussed in this paper, namely DUNE (left panel), JD (middle panel), and KD (right panel).
The cyan (black) ellipses in each panel correspond to the standard interaction case (obtained varying the CP phase $\delta_{\mathrm{CP}}$ in the range $[-180^{\circ},\, 180^{\circ}]$), referring to the  normal (inverted) mass hierarchy.
 The red (blue) colored blobs in the upper  panels show the bi-events in the presence of the non-unitarity parameters $|\alpha_{21}|$ with strength 0.025 for the NH (IH) case. The lower panels show the same but in the presence of $|\alpha_{31}|$ fixed to 0.1. In both cases, CP phase $\delta_{\mathrm{CP}}$ and the phase associated with corresponding off-diagonal NUNM parameter are varied in the range  $-180^{\circ}$ to $180^{\circ}$. 
The plots show that DUNE (left panels) is in principle expected to be able to distinguish the mass hierarchy even in presence of NUNM. On the other hand, the two standard model curves and the two shadowed regions overlap in the case of JD (center panels) and KD (right panels). For these, the amplitude of the shadowed regions when $\phi_{31}$ is varied is much smaller than the DUNE one since, as it is clear from the analytical formulas, $|\alpha_{31}|$ is always coupled to matter effects, which are smaller for the two J-PARC based experiments.

In our numerical simulations on the sensitivity to the NUNM parameters, the true values of the standard oscillation parameters have been chosen as in Table~\ref{table:vac}, while the true values of the NUNM parameters are set to zero. Since the presence of non-unitarity has basically the same effect on the number of events in the case of NH and IH, as shown in Fig.~\ref{fig:bi-event}, we consider only the NH case. The computation of the $\Delta\chi^2$
is based on the pull method~\cite{Huber:2002mx,Fogli:2002pt,Gonzalez-Garcia:2004pka} implemented in the GLoBES software.
We study the NUNM parameters by fixing the mixing angles $\theta_{12}$, $\theta_{13}$, and two mass-squared differences $\Delta m^2_{21}$, and $\Delta m^2_{31}$ both in data and theory at their best fit values as given in Table~\ref{table:vac}. 
We check that the marginalization over the atmospheric mass-squared difference $\Delta m^2_{31}$ does not have any significant effect on our analysis. On the other hand, the only notable effect of the marginalization over the reactor mixing angle $\theta_{13}$ (which has a very small experimental uncertainty of 3\%), is the worsening of the $\alpha_{11}$ bound at the level of 15\%. This is due to the fact that there is a correlation between these two parameters, which appear in a term proportional to $\alpha_{11}^2\sin^2 2\theta_{13}$ in the $\nu_\mu\to\nu_e$ transition probability as shown in Ref.~\cite{Escrihuela:2015wra}. 
Finally, we marginalized $\theta_{23}$ in its current 3$\sigma$ allowed range~\cite{Capozzi:2021fjo}, which is approximately [$40^{\circ},\, 50^{\circ}$] and the CP phase $\delta_{\mathrm{CP}}$ in its entire possible range $[-180^{\circ},\, 180^{\circ}]$. We keep both these parameters with true values as in Table~\ref{table:vac}. Moreover, we consider one NUNM parameter at a time, $\ie$,  when a parameter is taken into account the others are considered to be zero. In Appendix~\ref{sec:NU_marg}, we also discuss in detail how our results would be affected by the marginalization over multiple NUNM parameters. Finally, the phases associated with each off-diagonal NUNM parameters are marginalized over the entire possible range from $-180^{\circ}$ to $180^{\circ}$.

\section{Numerical results}
\label{results}

In this section, we provide a detailed discussion of the numerical results obtained with the GLoBES simulations. We study the correlations between the NUNM parameters and the two less constrained standard oscillation parameters $\theta_{23}$ and $\delta_{\mathrm{CP}}$. Then, we determine the expected bounds on all the non-standard parameters at 90\% confidence level (C.L.) when the expected data from DUNE, JD, KD, and their various combinations are taken into account.

\subsection{Correlations in test $(\theta_{23}-\alpha_{ij})$ and test $(\delta_{\mathrm{CP}}-\alpha_{ij})$ planes}
\label{subsec:correlation}
\begin{figure}[h!]
\centering
\includegraphics[width=\textwidth]{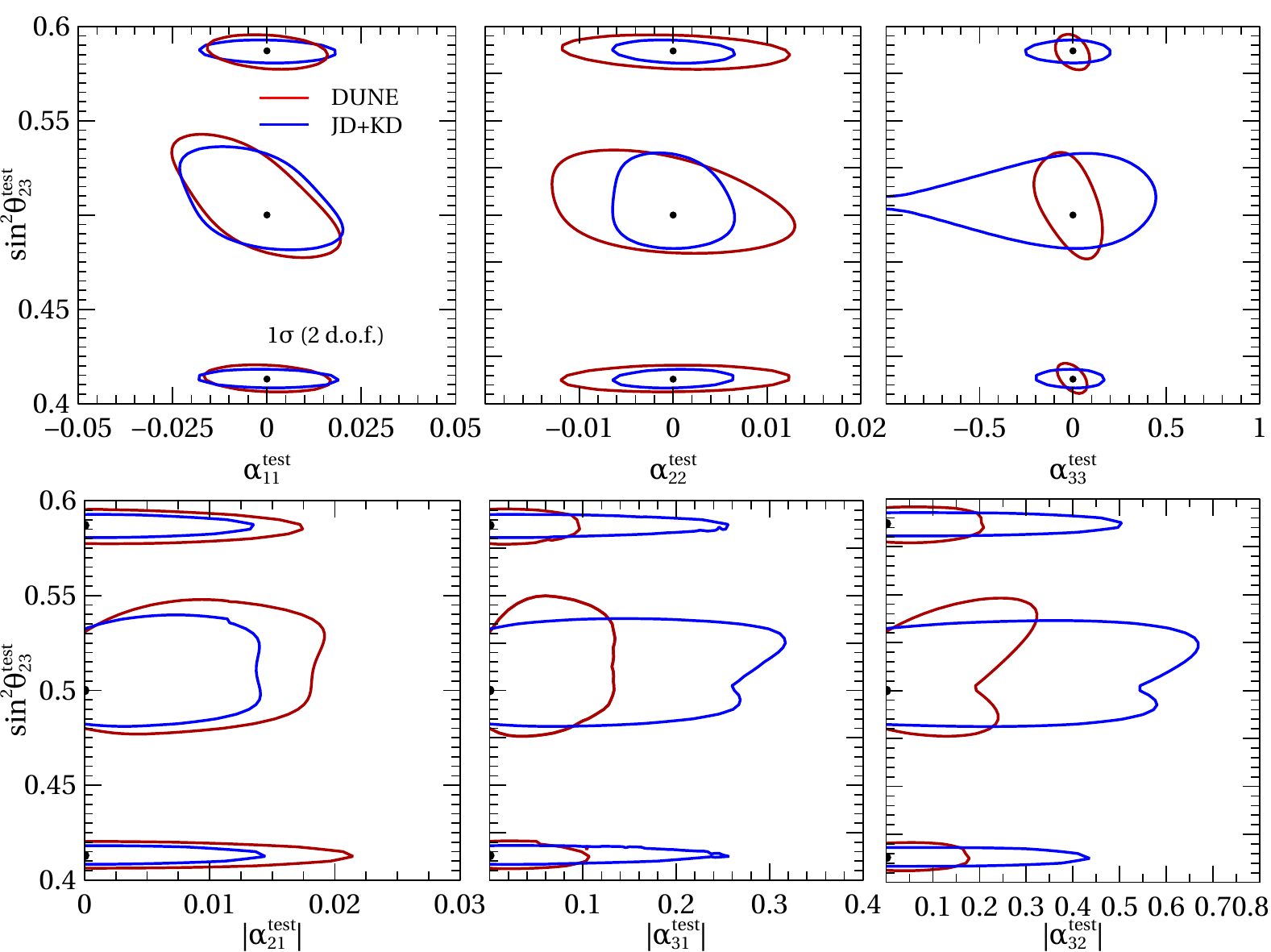}
\mycaption{1$\sigma$ (2 d.o.f.) C.L. contours in the $(\theta_{23}$ - $\alpha_{ij})$ planes for DUNE (red curves) and JD+KD (blue curves). The true values of $\theta_{23}$ considered here are $\theta_{23}=40^{\circ},\, 45^{\circ}$, and $50^{\circ}$ (shown by the black dots in each panel). True value of $\delta_{\mathrm{CP}}$ is considered to be $-\,90^{\circ}$. In the fit, $\delta_{\mathrm{CP}}$ and the phases associated with off-diagonal NUNM parameters (see lower panels) are marginalized over the range of [-$180^{\circ}$, $180^{\circ}$]. The other oscillation parameters are fixed at their best fit values (see Table~\ref{table:vac}).}
\label{fig:th23_corr}
\end{figure}

In Fig.~\ref{fig:th23_corr}, we show the correlation between various NUNM parameters and the atmospheric mixing angle $\theta_{23}$. 
 Three different true values of $\theta_{23}$ (shown by the black dot in each panel) have been analyzed, namely, $\theta_{23}=40^{\circ}$ ($\sin^2\theta_{23} = 0.413$) in the lower octant, the maximal value $\theta_{23}=45^{\circ}$ ($\sin^2\theta_{23} = 0.5$), and $\theta_{23}=50^{\circ}$ ($\sin^2\theta_{23} = 0.586$) in the upper octant. 
We keep the true value of $\delta_{\mathrm{CP}}$ as $-90^{\circ}$. Red contours correspond to 1$\sigma$ (2 d.o.f.) C.L. obtained using the DUNE setup, while the blue contours show the same for the  JD+KD combination. We observe that the correlation between $\sin^2\theta_{23}$ and $\alpha_{11}$ (top left panel) is almost the same for both DUNE and JD+KD setups. In the $(\theta_{23}-\alpha_{22})$ and $(\theta_{23}-\alpha_{21})$ planes, the allowed regions for JD+KD are smaller than the ones corresponding to DUNE, which suggest that the J-PARC based experiments will have a better sensitivity on these parameters (see the following subsection). For the other three NUNM parameters, DUNE shows much better sensitivity than JD+KD, due to the larger matter effects which couples to the NUNM parameters $\alpha_{33}$, $|\alpha_{31}|$, and $|\alpha_{32}|$. In particular, for maximal value of $\theta_{23}$, there are no lower limits on $\alpha_{33}$ from JD+KD. One interesting point to be noted is that, in the presence of non-unitary mixing, $\sin^2\theta_{23}$ can be constrained better for non-maximal true values of $\theta_{23}$.

\begin{figure}[tbp]
\centering
\includegraphics[width=\textwidth]{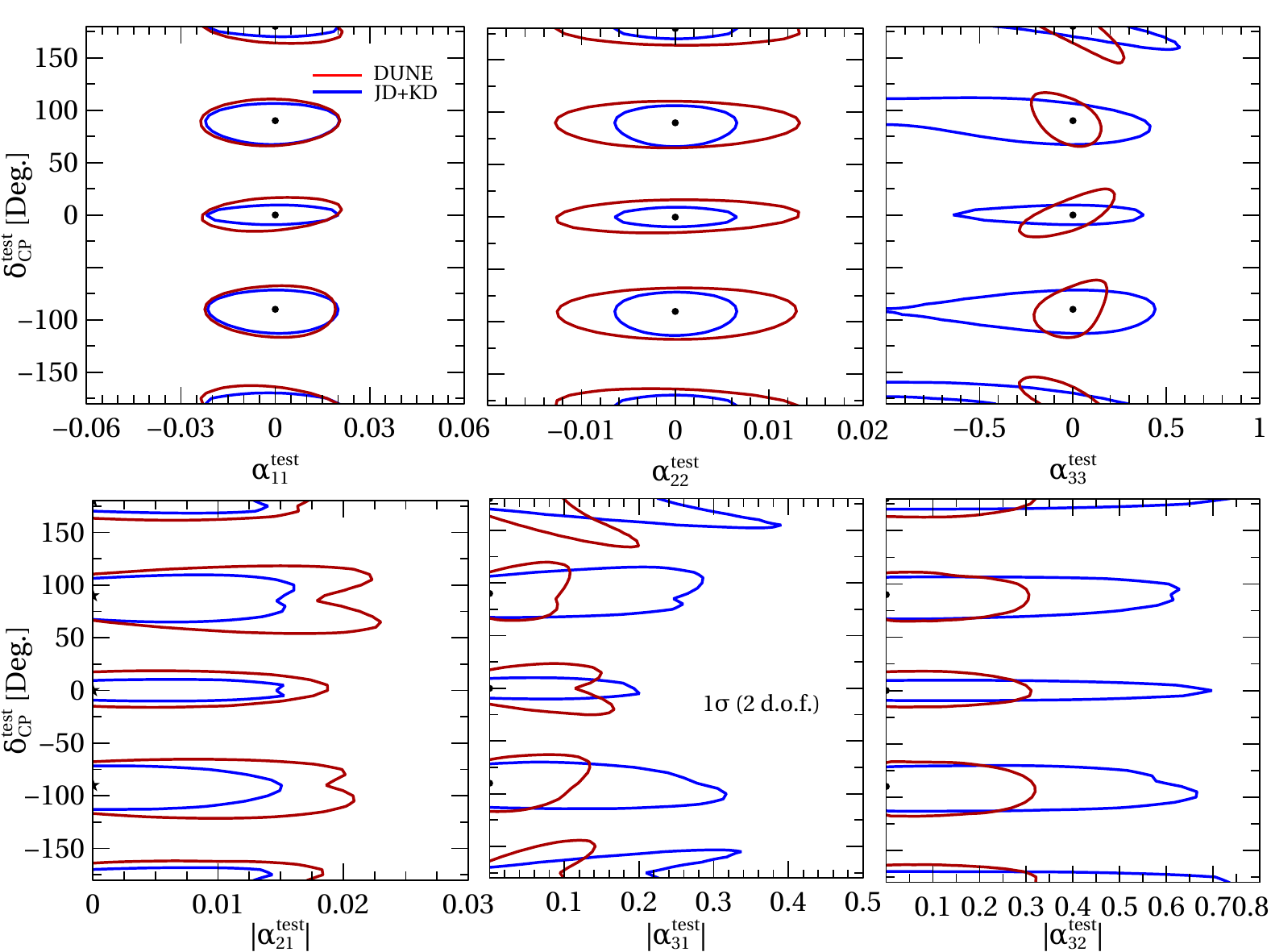}
\mycaption{
1$\sigma$ (2 d.o.f.) C.L. contours in the $(\delta_{\mathrm{CP}}$-$\alpha_{ij})$ planes for DUNE (red curves) and JD+KD (blue curves). The true values of $\delta_{\mathrm{CP}}$ considered here are $\delta_{\mathrm{CP}}=0^{\circ},\, 90^{\circ},\, 180^{\circ},\, -90^{\circ}$ (shown by the black dots in each panel). The true value of $\theta_{23}$ is $45^{\circ}$, while its test value is kept free in the range [$40^{\circ}$, $50^{\circ}$]. In the fit, the phases associated with off-diagonal NUNM parameters (see lower panels) are marginalized over the range of [-$180^{\circ}$, $180^{\circ}$]. The other oscillation parameters are fixed at the values as given in Table~\ref{table:vac}.
}
\label{fig:dcp_corr}
\end{figure}

In Fig.~\ref{fig:dcp_corr}, we show the correlation of various NUNM parameters with the CP phase $\delta_{\mathrm{CP}}$.
In each panel, the black dots correspond to the true values of $\delta_{\mathrm{CP}}=0^{\circ},\, 90^{\circ},\, 180^{\circ}$, and $-\,90^{\circ}$ and the benchmark value chosen for the $\alpha_{ij}$ under consideration. The red contours refer to the sensitivity at 1$\sigma$ (2 d.o.f.) C.L. expected to be obtained by DUNE setup and the blue curve shows the same for JD+KD combination. The plots follow a similar behavior already seen in Fig.~\ref{fig:th23_corr}. In particular, the allowed regions for the two experiments are almost the same in the case of $\alpha_{11}$, smaller for JD+KD in the case of $\alpha_{22}$ and $|\alpha_{21}|$ and, finally, much smaller for DUNE in the case of the NUNM parameters $\alpha_{33}$, $|\alpha_{31}|$, and $|\alpha_{32}|$. Note that when true $\delta_{\mathrm{CP}}$ = $90^{\circ}$, $180^{\circ}$, and $-\,90^{\circ}$, the J-PARC based detectors will not be able to set any lower limit on $\alpha_{33}$.

\subsection{Constraints on non-unitary neutrino mixing parameters}
\label{subsec:num_results}

In this subsection, we present our numerical results showing the expected constraints on the six NUNM parameters ($\alpha_{ij}$) that DUNE, JD, KD, and JD+KD setups can place. 
We derive bounds on $\alpha_{ij}$ following the simulation method as discussed in Sec.~\ref{sec:sim-details}.
The statistical significance with which we can constrain the NUNM parameters ($\alpha_{ij}$) in a given experiment is defined as 

\begin{equation}
	\Delta \chi^2 = \underset{(\theta_{23},\,\delta_{\mathrm{CP}},\,\phi_{ij},\,\lambda_1,\, \lambda_2)}{\mathrm{min}} \,\bigg[\chi^2(\alpha_{ij}\neq0)-\chi^2(\alpha_{ij}=0)\bigg]\,\, ,
\end{equation}
where, $\chi^{2} (\alpha_{ij}\neq0)$ and $\chi^2(\alpha_{ij}=0)$ are calculated by fitting the prospective data assuming NUNM ($\alpha_{ij}\neq0$) and UNM $(\alpha_{ij}=0)$. Note that $\chi^2(\alpha_{ij}=0) \approx 0$ because the statistical fluctuations are suppressed to obtain the median sensitivity of a given experiment in the frequentist approach~\cite{Blennow:2013oma}. While estimating the constraints, we marginalize over the most uncertain oscillation parameters ($\theta_{23}$, $\delta_{\mathrm{CP}}$) and the phases associated with the off-diagonal NUNM parameters ($\phi_{ij}$) in the fit. We also minimize over the systematic pulls on signal ($\lambda_1$) and background ($\lambda_2$).

\begin{figure}[h!]
\centering
\includegraphics[width=\textwidth]{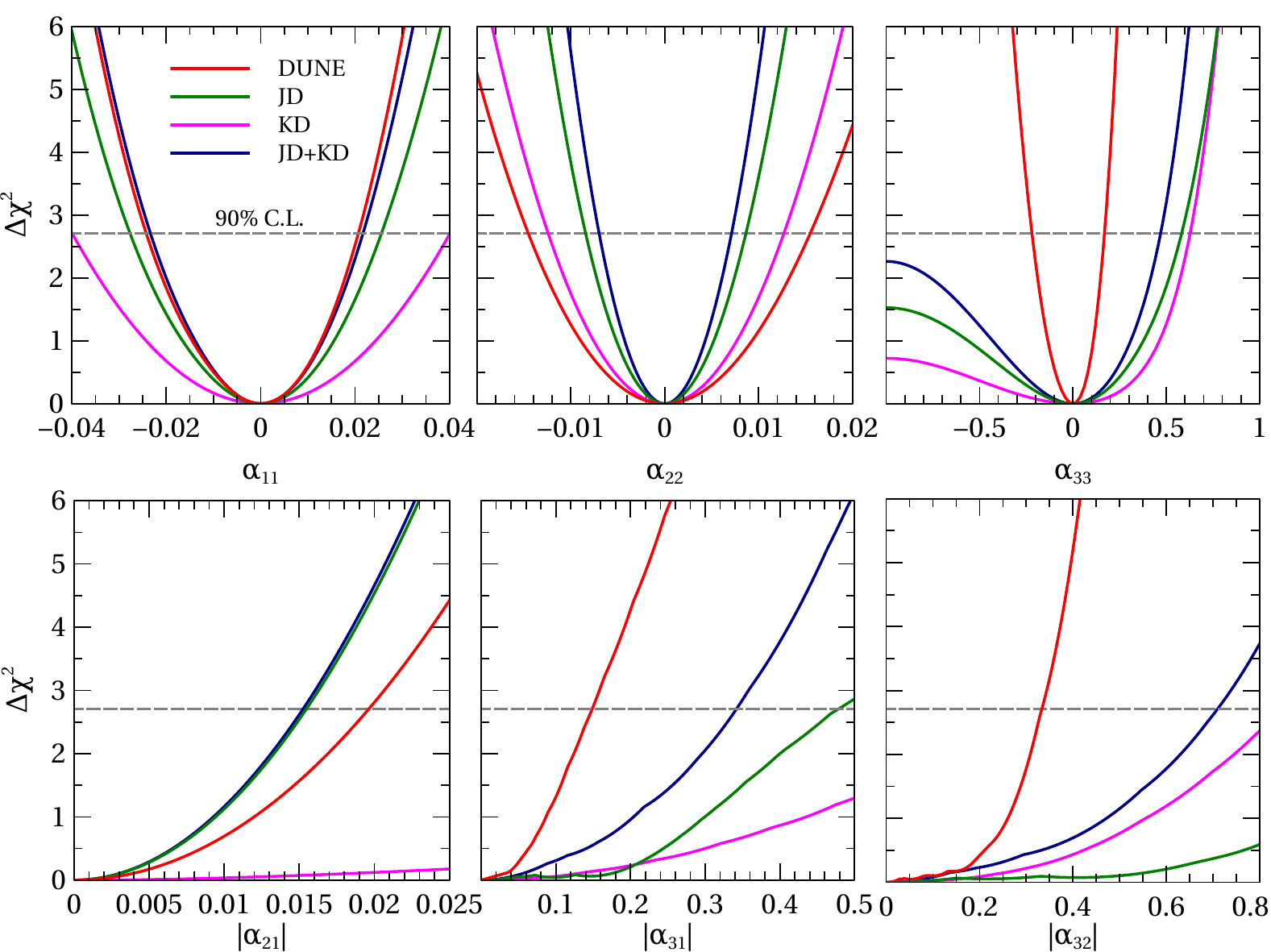}
\mycaption{ Expected limits on the NUNM parameters
	from DUNE (red curves), JD (green curves), KD (magenta curves), and JD+KD (blue curves).
	The upper (lower) panels are for the diagonal (off-diagonal) NUNM parameters one at a time.
	True values of $\theta_{23}$ and $\delta_{\mathrm{CP}}$ are $45^{\circ}$ and $-\,90^{\circ}$, respectively. 
	For the diagonal NUNM parameters, we marginalize over $\theta_{23}$ in the range [$40^{\circ},   50^{\circ}$] and $\delta_{\mathrm{CP}}$ in the range [$-180^{\circ}, 180^{\circ}$] in the fit. For the off-diagonal NUNM parameters, apart from $\theta_{23}$ and $\delta_{\mathrm{CP}}$, we also marginalize over the associated NUNM phases in the range of $-180^{\circ}$ to $180^{\circ}$.}
\label{fig:chi_sq}
\end{figure}

In Fig.~\ref{fig:chi_sq}, we plot the $\Delta \chi^2$ function for the six NUNM parameters analyzed in our paper considering only the far detectors in a given setup. Upper (Lower)  panels show the results for the diagonal (off-diagonal) NUNM parameters. The red curves in  each panel refer to the sensitivities obtained for the DUNE setup considering a total of $336$ kt-MW-yrs exposure, corresponding to a total 7 years of data collection with equal run-time in neutrino and antineutrino modes. The green curves show the results for JD  for which we consider a total exposure of 2431 kt-MW-yrs with 10 years of total run-time (2.5 years in neutrino mode and 7.5 years in antineutrino mode). The magenta curves correspond to KD assuming the same  exposure. We also estimate the results for JD+KD as shown by the blue curves.
 From the upper left panel, we observe that DUNE and JD+KD place similar constraints on $\aee$. The sensitivity to this parameter comes from two contributions: disappearance of intrinsic $\nu_e$ beam and the $\nu_e$ appearance.
 For both of these contributing channels, 
 DUNE has better systematics and JD+KD has more statistics. As a result, limits on $\alpha_{11}$ is found to be almost the same for the two setups. However, for $\amm$ (upper middle panel), JD+KD has significantly better sensitivity compared to DUNE setup. This is because $\amm$ is mainly constrained by the disappearance channels, which due to the large statistics, is primarily limited by the systematic uncertainties. Since the normalization error for this channel is 3.5\% (5\%) for JD+KD (DUNE), it is clear that JD+KD can put better limit than DUNE. We have checked that if we consider the same amount of systematic uncertainties for both the setups, DUNE shows slightly better sensitivity than T2HKK.
 
For $|\ame|$, DUNE and JD+KD have comparable sensitivities (see lower left panel). The slightly better limit on $|\alpha_{21}|$ achieved in the case of JD+KD as compared to DUNE, is due to the fact that JD+KD has larger statistics in the appearance channels. For the other three parameters $|\ate|$, $|\atm|$ and $\att$, which enter the $\nu_{\mu}\rightarrow\nu_{e}$ appearance channel through matter parameters $\Delta_{e}$ and $\Delta_n$ (see Eq.~\ref{eq:pme_anylit_main}), DUNE outperforms JD+KD setups because of its large matter effects.

\begin{table}[thb!]
\begin{center}
 \begin{adjustbox}{width=\textwidth}
 \begin{tabular}{|c|c|c|c|c|c|c|}
\hline\hline
&DUNE&JD&KD&JD+KD&JD+KD+DUNE&T2K+NO$\nu$A\\ 
\hline
$\alpha_{11}$ & [-0.020, 0.020] &[-0.025, 0.025]&[-0.040, 0.040]&   [-0.022, 0.022] & [-0.017, 0.017]& [-0.06, 0.06]\\
\hline
$\alpha_{22}$& [-0.014, 0.014]& [-0.0087, 0.0087]&[-0.013, 0.013]& [-0.007, 0.007] & [-0.006, 0.006] & [-0.02, 0.02]\\
\hline
$\alpha_{33}$& [-0.2, 0.17] & $<$ 0.6 & $<$ 0.63 & $<$ 0.476  & [-0.17, 0.17]& $<$ 0.64\\
\hline
$|\alpha_{21}|$ &  $<$ 0.022 & $<$ 0.015 & $<$ 0.10  & $<$ 0.016 & $<$ 0.012 & $<$ 0.06\\ 
\hline
$|\alpha_{31}|$& $<$ 0.15  & $<$ 0.48 & $<$ 0.70 & $<$ 0.34 &$<$ 0.11 & $<$ 2.20\\
\hline
$|\alpha_{32}|$& $<$ 0.33 & $<$ 1.2 & $<$ 0.85 & $<$ 0.71& $<$ 0.27 & $<$ 1.4\\
\hline\hline
\end{tabular}
 \end{adjustbox}
\mycaption{Bounds on the NUNM parameters at 90\% C.L. (1 d.o.f.) using DUNE (second column), JD (third column), KD (fourth column), and JD+KD (fifth column). Sixth column shows the results for the combination of DUNE and JD+KD. Last column depicts results using the full exposure of T2K and NO$\nu$A.}
\label{Tab:constraints}
\end{center}
\end{table}

We summarize our results in Table~\ref{Tab:constraints}, where we give the bounds on the six NUNM parameters at 90\% C.L. for the various long-baseline  experimental setups discussed in this paper. As clear from our previous discussion,  the expected constraints on NUNM parameters from DUNE is better than the other two experiments JD and KD (and their combination) except for the parameters $\amm$ and $|\ame|$, where JD has better sensitivity than DUNE.
Finally, in the sixth column of the Table, we give the final constraints on the NUNM parameters by combining the expected results from  DUNE and JD+KD setups. As we have anticipated, the bounds experience a general improvements by $\sim\,20$\%, with the precise magnitude depending on the parameter under consideration.

For a comparison with the ongoing long-baseline experiments, we also add the expected constraints from the combination of the T2K and NO$\nu$A setup in the last column. For T2K, we consider a total exposure of 84.4 kt-MW-yrs with 22.5 kt detector mass, 750 kW proton beam power, 5 years run-time (2.5 years each for neutrino and antineutrino mode). 
 For NO$\nu$A, the considered exposure is 58.8 kt-MW-yrs with 14 kt detector mass, 700 kW proton beam power, and 6 years for total run-time (3 years each for neutrino and antineutrino modes).
Due to the limited statistics, we observe that the expected constraints from T2K+NO$\nu$A setup is worse than the DUNE or JD+KD setup.

\begin{table}
	\centering
	\begin{tabular}{|c|c|c|}
		\hline\hline
		Parameter & DUNE (3.5 yrs+3.5 yrs)& DUNE (5 yrs+5 yrs)\\ 
		\hline
		$\alpha_{11}$& [-0.020, 0.020] & [-0.020, 0.018] \\
		\hline
		$\alpha_{22}$ & [-0.014, 0.014] & [-0.013, 0.013]\\
		\hline
		$\alpha_{33}$ & [-0.2, 0.17] & [-0.19, 0.15]\\
		\hline
		$|\alpha_{21}|$ & $<$ 0.022 & $<$ 0.016\\ 
		\hline
		$|\alpha_{31}|$ & $<$ 0.15 & $<$ 0.12\\
		\hline
		$|\alpha_{32}|$ & $<$ 0.33 & $<$ 0.31\\
		\hline\hline
	\end{tabular}
	\mycaption{
		90\% C.L. (1 d.o.f.) limits on  the NUNM parameters considering two different exposures of DUNE: total run-time of 7 years (see second column) and 10 years (see third column) equally divided in neutrino and antineutrino modes.}
	\label{tab:DUNE_runtime}
\end{table}

Note that if some information coming from the near detector (for an example, measurement of the initial neutrino flux) are used to analyze the far detector data then the constraints on $\alpha_{11}$ and $\alpha_{22}$ may be modified (see Sec. \ref{sec:ND} for a detailed discussion). 
	However, in this section, we adopt a different strategy, where we simulate the far detector data alone to set limits on the NUNM parameters. In principle, this approach is valid if the initial flux can be predicted by some theoretical calculation or measured at an experiment which is insensitive to neutrino oscillation phenomena. In fact, the MINOS/MINOS+ experiment adopted this strategy where the oscillation data at the far detector was analyzed using information from the MINERvA flux predictions~\cite{MINERvA:2016iqn}. In future, if somehow we can apply this approach for DUNE and T2HKK, then in principle, one can estimate the limits on the NUNM parameters using only the far detector data.

	At the same time, we understand that the assumptions that we take in our paper for the systematic uncertainties at the far detector may be too optimistic if we do not use the near detector to measure the initial flux.
		To have a better understanding on this issue, we perform some study and find that limits on $\alpha_{11}$ and $\alpha_{22}$ are mainly governed by the systematic uncertainties. In other words, the expected constraints on these two NUNM parameters are proportional to the systematic uncertainties that we consider in our simulation. Therefore, it may be possible to predict what would be the limits on $\alpha_{11}$ and $\alpha_{22}$ for a given assumptions on systematic uncertainties.

We compare our results summarized in Table~\ref{Tab:constraints} with the bounds reported in Table~\ref{alphabounds1}\footnote{In order to get their results, the authors of Ref.~\cite{Escrihuela:2016ube} left free the standard oscillation parameters $\theta_{23}$, $\delta_{\mathrm{CP}}$, and $\Delta m^2_{31}$ and the NUNM parameters $\aee$, $\ame$, $\amm$. Conversely, in our work we marginalize over $\delta_{\mathrm{CP}}$ and $\theta_{23}$ only, but we have checked that the marginalization over $\Delta m^2_{31}$ does not have any significant impact. In Appendix~\ref{sec:NU_marg}, we also show that the marginalization over $\aee$, $\ame$, and $\amm$ does not cause remarkable change in the results.}. We observe that the bound we achieve from the DUNE+JD+KD (DUNE+T2HKK) setup for the diagonal $\aee$ is $\sim$\,80\% better than the bound quoted in Ref.~\cite{Forero:2021azc}. In the $\amm$ case, the two results are comparable, with a slightly better limit when the global neutrino data analysis is performed. 
For the remaining diagonal parameter $\att$, NC data from MINOS/MINOS+ give a 60\% stronger bound \cite{Forero:2021azc}  compared to the one expected from the DUNE+JD+KD setup. As for $\ame$, the authors of the Ref.~\cite{Forero:2021azc} make use of the triangular inequality as well as the data from the short-baseline experiments; this allows to constrain the mentioned parameter very tightly. However, due to the large statistics and good systematics of DUNE and JD+KD setups, we can achieve an almost similar bound without using any external hypothesis on the relations between the $\alpha_{ij}$. On the other hand,
constraints on $\ate$ and $\atm$ in Table~\ref{alphabounds1} are substantially better than the ones we obtain from DUNE+JD+KD setup. Also, in these cases, the triangular inequalities which link them to the diagonal NUNM parameters  play an important role, together with the short-baseline experiments limits on the $\nu_\tau$ appearance. However, it is important to stress that all our results are obtained in a complete model independent fashion, relying only on the expected data from DUNE and T2HKK. We check that for our best setup, namely DUNE+T2HKK, the only parameter whose limit gets improved when we consider these inequalities is $|\alpha_{32}|$ because of the stringent bound on $\alpha_{22}$. Since $|\alpha_{21}|$ is already tightly constrained, we do not see any improvement in its limit using these inequalities. As far as the bound on $|\alpha_{31}|$ is concerned, since the limit  on the diagonal parameter $\alpha_{33}$ is very poor, we also do not see any improvement.

Recently, the DUNE collaboration~\cite{DUNE:2021cuw}  exploited the possibility of increasing the exposure of the experiment from 336 kt-MW-yrs to 480 kt-MW-yrs (corresponding to an increase of the  data taking time  from 7 years to 10 years with 5 years in neutrino mode and 5 years in antineutrino mode). In Table~\ref{tab:DUNE_runtime}, we compare our previous constraints from the DUNE experiment, Table~\ref{Tab:constraints}, with those obtained in the $(5 + 5)$ years configuration. We observe that the constraints on  all six NUNM parameters improve by small amount except for $|\ame|$, which shows a significant improvement. This happens because the higher run-time increases statistics of the $\nu_{\mu}\rightarrow\nu_e$ appearance channel, which is the one driving the $\alpha_{21}$ sensitivity. On the other hand, the $\nu_\mu\rightarrow\nu_{\mu}$ disappearance channel is almost already saturated by systematics after 3.5 years + 3.5 years of running. This leads to only small improvements on the other NUNM parameters sensitivities.

\section{Benefits of having near detectors}
\label{sec:ND}

Near detectors (ND) are a fundamental component for long-baseline neutrino experiments. Indeed, a detector placed very close to the beam source (from hundreds of meters to a few kilometers) is able to monitor the neutrino beam, measuring the flavor composition, and the total number of neutrinos emitted from the source. 
Near detectors are not expected to improve any of the standard oscillation parameter measurements, since at such short distances, oscillations do not develop for neutrinos with energies in the GeV range.
 However, in some new physics scenarios, in which, oscillation probabilities contain  zero-distance terms, near detectors can be used to constrain non-standard parameters in a very straightforward way. This is the case of the non-unitarity framework under discussion where, as already mentioned in Sec.~\ref{formulae}, at vanishing distances we have zero-distance terms in case $\nu_{\mu}\rightarrow\nu_{e}$ appearance channel: $
P_{\mu e}^{L = 0} \sim |\alpha_{21}|^2$, and $\nu_{\mu}\rightarrow\nu_{\mu}$ disappearance channel:
$P_{\mu\mu}^{L = 0} \sim 1+ 2 |\alpha_{21}|^2+6 \alpha_{22}^2+4 \alpha_{22}\,.$
Thus, we can expect that T2HKK and DUNE near detectors would be able to constrain two parameters $|\alpha_{21}|$ and $\alpha_{22}$ from $\nu_{\mu}\rightarrow\nu_e$ appearance and $\nu_{\mu}\rightarrow\nu_{\mu}$ disappearance channel, respectively, but also $\alpha_{11}$ considering the $\nu_e$ beam contamination (see Eq.~\ref{eq:e-disapp}). So, in this section, we analytically infer the order of magnitude of bounds implied by ND measurements.
Let us  consider the total number of events of a given channel as~\cite{Giarnetti:2020bmf}:
\begin{equation}
    N= N_0 P_{\alpha\beta}(\alpha_{ij})\,,
\end{equation}
where, the normalization factor $N_0$ includes all the detector properties. For an oscillation channel $\nu_\alpha\to\nu_\beta$, $N_0$ can be defined as:
\begin{eqnarray}
\label{eq:asirate2}
N_0 &=&\int_{E_\nu} dE_\nu \,\sigma_\beta(E_\nu)\,\frac{d\phi_\alpha}{dE_\nu}(E_\nu) \,\varepsilon_\beta(E_\nu)\,,
\end{eqnarray} 
where, $\sigma_\beta$ denotes the production cross-section of  the $\beta$ lepton, $\varepsilon_\beta$ represents the detector efficiency, and $\phi_\alpha$ stands for the initial neutrino flux of flavor $\alpha$. If we want to put bounds on new physics parameters, we can use a simple $\chi^2$ test with a gaussian $\chi^2$ defined as 
\begin{equation}
    \chi^2=\frac{(N_{obs}-N_{fit})^2}{\sigma^2}\,,
\end{equation} 
where, $\sigma$ represents the uncertainty on the number of events; in this case, neglecting the backgrounds, we get:
\begin{equation}
    \chi^2=\frac{N_0^2}{\sigma^2}\left[\delta_{\alpha\beta}- P_{\alpha\beta}(\alpha_{ij}^{fit})\right]^2\,.
\label{chi}
\end{equation}
For the $\nu_{\mu}\rightarrow\nu_{\mu}$ disappearance channel, the leading term of the probability is $P_{\mu\mu}^{L=0}=1+4\alpha_{22}$. Therefore, the $\chi^2$ assumes the form:
\begin{equation}
    \chi^2=\frac{16 N_0^2 \alpha_{22}^2}{\sigma^2}.
\end{equation}
At a chosen confidence level, represented by a cut at $\chi^2_{0}$, it is possible to exclude the region satisfying:
\begin{eqnarray}
|\alpha_{22}|>\frac{\sqrt{\chi^2_0}\sigma}{4 N_0}.
\label{chi_dis}
\end{eqnarray}

\begin{table}
	
	\centering
	
	\begin{tabular}{|cc|*{4}{c|}}
		
		\hline\hline
		
		\multicolumn{2}{|c|}{\multirow{2}{*}{}} & \multicolumn{2}{|c}{$\alpha_{11}$} & \multicolumn{2}{|c|}{$\alpha_{22}$}\\
		
		\cline{3-6}
		
		\multicolumn{2}{|c|}{Expt.} & w/o norm. & w/ norm. & w/o norm. & w/ norm. \\
		
		\hline\hline
		
		\multicolumn{2}{|c|}{DUNE} & [-0.02, 0.02] & [-0.043, 0.034] &  $[-0.014, 0.014]$& [-0.036, 0.048]   \\
		
		
		\hline\hline
		
		\multicolumn{2}{|c|}{JD+KD} &  $[-0.022, 0.022]$ & [-0.048, 0.040]  &  $[-0.007, 0.007]$& [-0.038, 0.050] \\
		\hline\hline
		
		\multicolumn{2}{|c|}{DUNE+JD+KD} & $[-0.017, 0.017]$ & [-0.036, 0.026]  &  $[-0.006, 0.006]$& [-0.026, 0.039]   \\
		\hline

		\hline
		\hline
		
	\end{tabular}
	\mycaption{90\% C.L. (1 d.o.f.) limits on the NUNM parameters $\aee$ and $\amm$ for the two setups, DUNE, JD+KD, and combination of them. Second  and fourth column correspond to the constraints assuming only far detector. Third and fifth columns correspond to the constraints using the FD and ND correlation (or with normalization factor in the oscillation probability). }
	\label{Table:cons_w_norm}
\end{table}

\begin{figure}[h!]
\centering
\includegraphics[width=\textwidth]{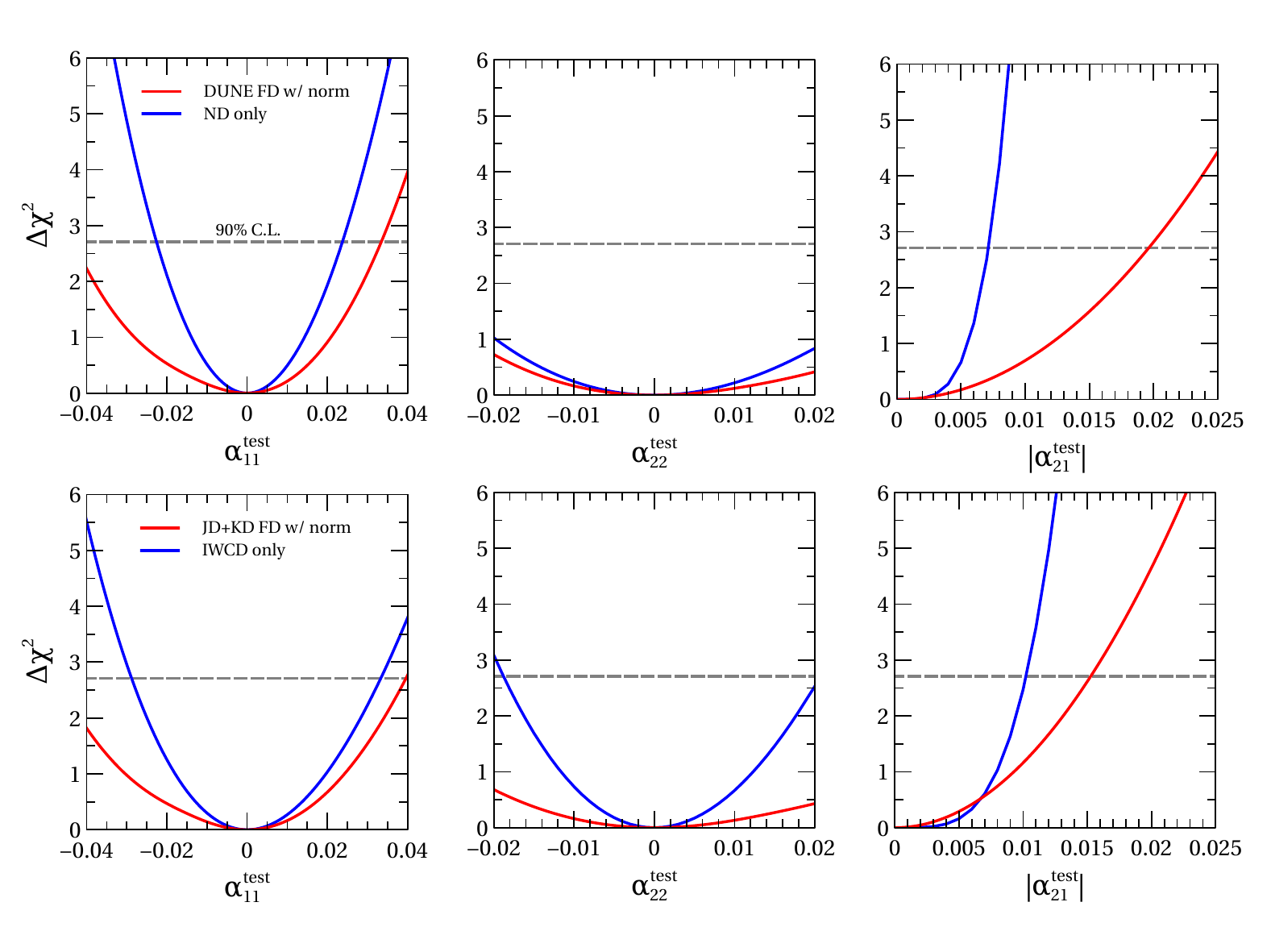}
\mycaption{Upper panels show the improvements in the sensitivities to $\alpha_{11}$, $\alpha_{22}$, and $|\alpha_{21}|$ due to the presence of 67 tons LArTPC near detector placed at distance of 574 meters from the neutrino source for DUNE. Lower panels portray the same for JD+KD having a 1 kt water Cherenkov  near detector placed at a distance of 1 km from J-PARC. The blue curves show the performance with only near detectors. The red curves represent the combined sensitivities due to both near and far detectors. True values of the standard oscillation parameters are taken from Table~\ref{table:vac}.
We obtain our results marginalizing over $\delta_{\mathrm{CP}}$ in the range $[-180^{\circ},\,180^{\circ}]$ and $\theta_{23}$ in the range $[40^{\circ},\,50^{\circ}]$ in the fit. We also marginalize over the associated phase $\phi_{21}$ in the range $[-180^{\circ},\,180^{\circ}]$ for the off-diagonal NUNM parameter $|\alpha_{21}|$.}
\label{fig:ND-DUNE}
\end{figure}

Since the disappearance channel is expected to produce a huge number of events at the near detector, one can consider the uncertainty to be dominated by systematic errors $\sigma_{sys}$. Thus, it is possible to approximate $\sigma\,\sim\,N_0 \sigma_{sys}$, where $N_0$ represents the number of events in absence of zero-distance effects, being the disappearance probability in that case equal to 1. This allows to simplify  Eq.~\ref{chi_dis} as follows:
\begin{eqnarray}
|\alpha_{22}|>\frac{\sqrt{\chi^2_0}\sigma_{sys}}{4}\,,
\end{eqnarray}
which tells us that, neglecting backgrounds effects, the near detector limits would be of the order of the chosen systematic uncertainty. A similar approach can be used for the $\nu_e\to\nu_e$ oscillation channel (see Eq.~\ref{eq:e-disapp}), which arises from the $\nu_e$ beam  contamination, obtaining an inequality for $|\alpha_{11}|$ of the similar form:
\begin{eqnarray}
|\alpha_{11}|>\frac{\sqrt{\chi^2_0}\sigma_{sys}^{\nu_e}}{4}\,,
\label{a11_ND}
\end{eqnarray}
where $\sigma_{sys}^{\nu_e}$ refers to the systematic uncertainty on the $\nu_e\to\nu_e$ transition.\\
For the appearance channel, the zero-distance probability reads $P_{\mu e}^{L=0}=|\alpha_{21}|^2$; the $\chi^2$ function can therefore be written as:
\begin{equation}
    \chi^2=\frac{N_0^2 |\alpha_{21}|^4}{\sigma^2}\,,
\end{equation}
and the excluded region is expected to be determined by the following relation:
\begin{eqnarray}
|\alpha_{21}|<\sqrt[4]{\frac{\chi_0^2 \sigma^2}{N_0^2}}\,.
\end{eqnarray}
Since the number of events at the near detector is in principle very small (being only caused by new physics), the uncertainty is dominated by statistics. Thus, given a certain number of observed events, $\sigma\,\sim\,\sqrt{N_{obs}}$ and the excluded values of $|\alpha_{21}|$ reduced to:
\begin{eqnarray}
 |\alpha_{21}|<\sqrt[4]{\frac{\chi_0^2 N_{obs}}{N_0^2}}\,,
 \label{mue_ND}
\end{eqnarray}
suggesting that the bounds are very sensitive to the number of events and to the  running time of the experiment. 

Both DUNE and T2HKK will have near detectors~\cite{Miranda:2018yym,DUNE:2021tad,Coloma:2021uhq} which may play a crucial role to probe various new physics scenarios including the possibility of non-unitarity of the PMNS matrix which is the main thrust of this work. In our analysis, for DUNE, we consider a 67 tons LArTPC near detector placed at a baseline of 574 meters from Fermilab~\cite{DUNE:2021tad}. For JD+KD, we consider a 1 kt water Cherenkov near detector located at a baseline of 1 km from J-PARC which is known as IWCD~\cite{Drakopoulou:2017qdu, Wilson:2020trq}.
In order to simulate their responses, we scale the far detector fluxes for ND baselines and take into account their fiducial masses. We follow a very conservative approach as far as the systematic uncertainties at the near detectors are concerned. We multiply the FD systematic uncertainties by a factor of three and consider them as inputs for the ND.
 In DUNE near detector, we expect $\mathcal{O}(10^7)$ $\nu_\mu$ and $\bar{\nu}_{\mu}$ events, which provide bounds on $\alpha_{22}$. DUNE can place stringent constraints on $\alpha_{11}$ and $\alpha_{21}$ using $\mathcal{O}(10^6)$ $\nu_e$ and $\bar{\nu}_e$ events at ND, which stem from both intrinsic $\nu_e$ ($\bar{\nu}_e$) beam contamination and via $\nu_\mu\to\nu_e$ ($\bar{\nu}_\mu\rightarrow\bar{\nu}_e$) appearance caused due to zero-distance effect. 
  For the NDs, we consider their appropriate baselines, fiducial masses, and systematic uncertainties which we assume to be larger than the systematic uncertainties considered for the FDs. 
 
 \begin{table}
 	\centering
 	\begin{adjustbox}{width=\textwidth}
 		\begin{tabular}{|c|c|c|c|c|}
 			\hline\hline
 			Parameter & DUNE (ND) & DUNE FD w/ norm. &  IWCD & JD+KD w/ norm.\\ 
 			\hline
 			$\alpha_{11}$& [-0.020, 0.024] & [-0.043, 0.034] & [-0.029, 0.033] & [-0.048, 0.040] \\
 			\hline
 			$\alpha_{22}$ & [-0.033, 0.037] & [-0.036, 0.048] & [-0.019, 0.020] & [-0.038, 0.050] \\
 			\hline
 			$|\alpha_{21}|$ & $<$ 0.007 & $<$ 0.022 & $<$ 0.01 & $<$ 0.015 \\
 			\hline\hline
 		\end{tabular}
 	\end{adjustbox}
 	\mycaption{90\% C.L. (1 d.o.f.)	bounds on the NUNM parameters $\aee$, $\amm$, and $|\ame|$ obtained with and without near detectors in DUNE and JD+KD. Note that IWCD is the near detector for JD+KD setup.}
 	\label{tab:NearDetector}
 \end{table}
 
 Before discussing the limits that the near detectors would be able to set using their own data, we want to study the effect of the ND flux measurements on the far detector constraints. Indeed, if the initial neutrino flux is measured at the near detector and then extrapolated to the far detector, the probability which could be inferred at the far detector is the effective probability defined as
	\begin{equation}
		P^{\text{eff}}_{\alpha\beta} = \frac{P_{\alpha\beta}}{P^{L=0}_{\alpha\alpha}}\,.
		\label{eq:eff_prob}
	\end{equation}
The $P^{L=0}_{\alpha\alpha}$ term that appears in the denominator is the survival probability initial neutrino flavor at the source or the zero-distance term which act as a normalization factor. If we normalize the $\nu_\mu\to\nu_\mu$ survival probability at the far detector using the zero-distance term in Eq.~\ref{pmmzero}, it is observed that the contribution from $\alpha_{22}$ gets canceled at the leading order. As a result, sensitivity to the parameter $\alpha_{22}$ is worsened for a given setup.
The same happens for $\alpha_{11}$, whose contribution in the effective $\nu_e\to\nu_e$ disappearance probability is canceled at the leading order (see Eq.~\ref{eq:e-disapp} and~\ref{eq:zero-distance}). Since the sensitivity to this parameter arises partially due to the intrinsic $\nu_e$ that we have in the beam to begin with, the near detector normalization causes a deterioration of $\alpha_{11}$ limits.
In Table~\ref{Table:cons_w_norm}, we show how the constraints on $\aee$ and $\amm$ would be modified when taking into account the FD and ND correlation for the three setups namely, DUNE, JD+KD, and DUNE+JD+KD. We observe that the bound on $\aee$ is increased by a factor of almost two when we consider the correlation between the FD and ND. For $\amm$, the bound is deteriorated at least three times compared to the FD case only. We have checked that no other NUNM parameter is affected significantly if we consider the FD and ND correlation. Indeed, the non-diagonal parameters and $\alpha_{33}$ can be constrained using appearance channels (for which we do not have full cancellations in the effective probabilities) or using the interplay with matter effects, which are not developed at the near site. 
 
The bounds obtained using the above-mentioned near detectors are shown in Fig.~\ref{fig:ND-DUNE} for DUNE and T2HKK together with the results we got using the FD data with effective probabilities. For $\alpha_{11}$, NDs of the two setups can put bounds better than the one set by the FDs considering the ND normalization due to the very high statistics and the strong $\alpha_{11}$ dependence of the zero-distance probability. The improvement is roughly a factor of two for DUNE and 60\% for T2HKK (see Table~\ref{tab:NearDetector}). Note that the obtained limits are in agreement with the predictions deduced from Eq.~\ref{a11_ND}, once we insert a normalization uncertainty of 6\% (15\%) for DUNE (JD+KD).  

For the second diagonal parameter $\alpha_{22}$ we also observe a similar situation, in which the ND alone can put more stringent bounds than the far detector when the normalization is considered, despite of the increased systematics. The improvement can be quantified as roughly 25\% in DUNE and a factor of two in T2HKK. Once again, the analytical predictions from Eq.~\ref{a11_ND} are sufficiently recovered by the numerical simulations, considering that the near detectors normalization systematics are 15\% for DUNE and 10.5\% for T2HKK. Note that the bounds from the far detector data alone would be considerably better than the near detector ones. 

Finally, for the NUNM parameter $|\alpha_{21}|$ the near detectors bounds are considerably better than the far detector ones, due to the zero-distance effect outlined in Eq.~\ref{zerodistancemue}. 
In particular, the limits are $\sim$ 3 times smaller than the one set by the far detector in the DUNE facility and $\sim$ 70\% smaller in the case of T2HKK. Considering a number of observed events of  ${\cal O}(10)$, and taking into account that we expect $N_0\sim10^6$ per year \cite{DUNE:2021tad}, our analytic estimate  for $|\alpha_{21}|$ is comparable with the numerical results.

\section{Improvement due to the $\nu_{\tau}$ sample in DUNE}
\label{sec:tau-analysis}

The $\nu_\tau$ production at accelerator experiments is very challenging since the charged current interactions of such particles with nuclei have an energy threshold of 3.1 GeV. Thus, many proposed long-baseline experiments are not able to detect such neutrinos\footnote{In this respect, the OPERA setup provided neutrinos with an average  energy close to 13 GeV~\cite{OPERA:2010pne}.}.
However, the DUNE neutrino spectra will have peak at around 2.5 GeV (differently from the T2HKK where $E_\nu \simeq 0.6$ GeV) and the most energetic neutrinos of the beam will have enough energy to produce $\tau$ leptons. Recently, different studies~\cite{Machado:2020yxl,Ghoshal:2019pab,DeGouvea:2019kea,Martinez-Soler:2021sir} take into account the possibility of including the $\nu_\tau$ ($\bar{\nu}_\tau$) sample in the DUNE analysis. 

The recognition of such events could be in principle possible due to the imaging capabilities of LArTPC detectors. Because of a relatively small number of neutrinos with an energy above the production threshold and of the short lifetime of the $\tau$ leptons which could make the recognition of the $\tau$ interaction and the decay vertices a difficult task, the $\nu_\tau$ appearance channel is not really useful to constrain the standard oscillation parameters. However, when new physics affects the oscillation probabilities, it has been shown in Refs.~\cite{Ghoshal:2019pab,DeGouvea:2019kea,Denton:2021mso,Denton:2021rsa} that the NUNM parameters $\alpha_{33}$ and $|\alpha_{32}|$ constraints can be improved even by the small number of $\nu_\tau$ events. 

\begin{figure}[h!]
\centering
\includegraphics[width=\textwidth]{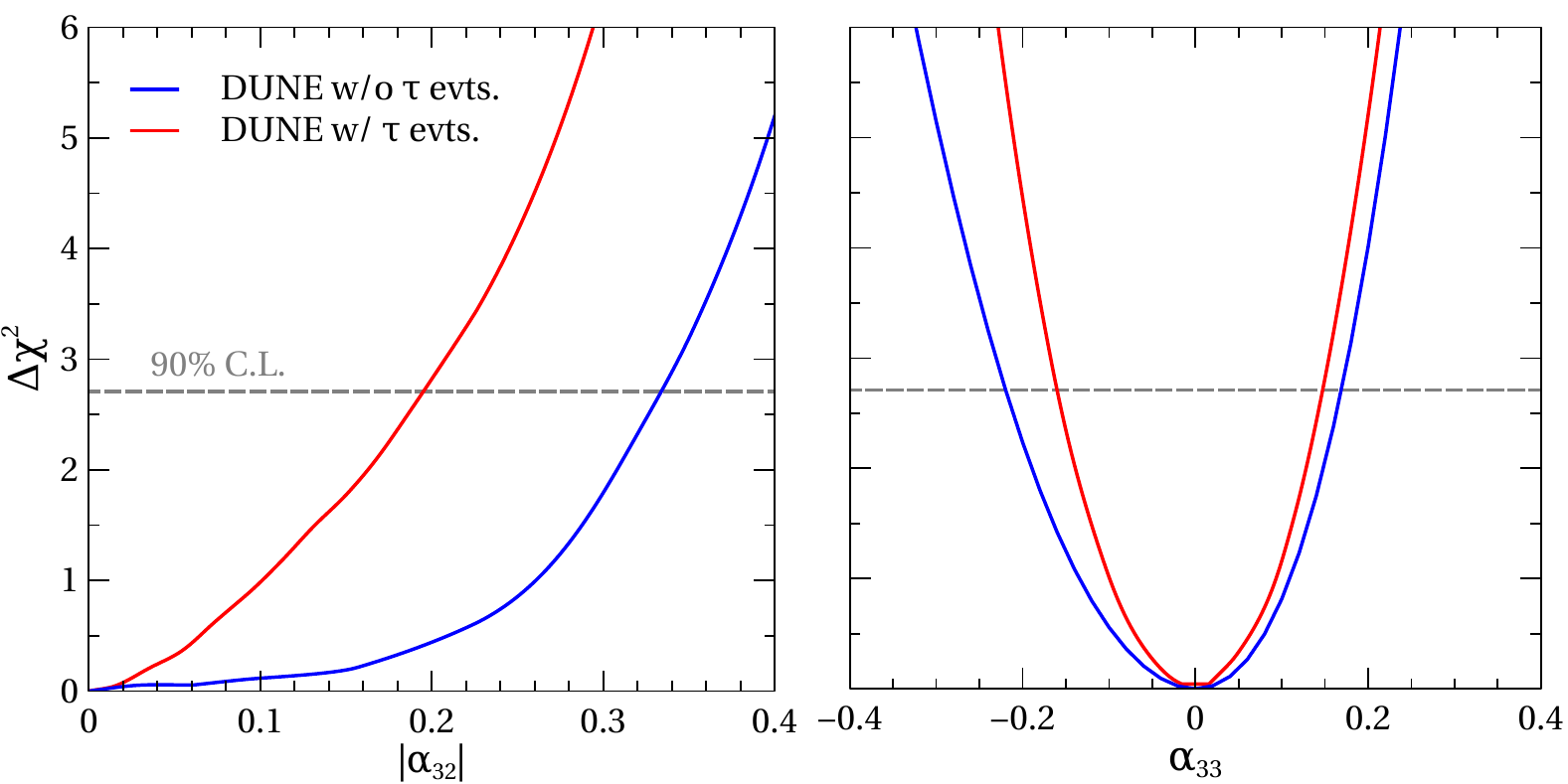}
\mycaption{Comparison between the DUNE sensitivities on $|\alpha_{32}|$ (left panel) and $\alpha_{33}$ (right panel) when $\nu_\tau$ appearance channel is included in the analysis (red lines) and the case where no $\tau$ events are analyzed (blue lines). True values of the standard oscillation parameters are taken from Table~\ref{table:vac}. All results have been obtained marginalizing over $\delta_{\mathrm{CP}}$ in the range $[-180^{\circ}, 180^{\circ}]$ and $\theta_{23}$ in the range $[40^{\circ}, 50^{\circ}]$. For $|\alpha_{32}|$ (left panel), we also marginalize over $\phi_{32}$ in the range $[-180^{\circ}, 180^{\circ}]$. }
\label{fig:tau}
\end{figure}

\begin{table}
	\centering
	\begin{tabular}{|c|c|c|}
		\hline\hline
		Parameter & w/o $\nu_{\tau}$ appearance & w/ $\nu_\tau$ appearance\\ 
		\hline
		$\alpha_{33}$ & [-0.2, 0.17] & [-0.16, 0.15] \\
		\hline
		$|\alpha_{32}|$ & $<$ 0.33 & $<$ 0.19 \\
		\hline\hline
	\end{tabular}
	\mycaption{90\% C.L. limits on the NUNM parameters $\alpha_{33}$ and $|\alpha_{32}|$ from the DUNE setup.
		Second (third) column shows the results without (with) $\tau$ in the analysis.}	
		\label{tab:DUNE_taus}
\end{table}
In case of non-unitary neutrino mixing, the $\nu_\mu\to\nu_\tau$ oscillation probability mainly depends on the three parameters $|\alpha_{32}|$, $\alpha_{33}$ and $\alpha_{22}$ (see Eq.~\ref{tauprob} in Appendix~\ref{Appendix:A}). While we expect that the sensitivity to $\alpha_{22}$ will not be improved by the small number of $\nu_\tau$ events, the other two poorly constrained NUNM parameters $|\alpha_{32}|$ and $\alpha_{33}$ could take advantage of $\nu_\mu\to\nu_\tau$ oscillation channel. We include $\tau$ events in our analysis in the following fashion.
\begin{itemize}
    \item For the hadronic decays of $\tau$ events having a branching ratio of 65\%, a 30\% signal efficiency has been assumed and 10\% of the NC events are considered as background~\cite{DeGouvea:2019kea}.
   
    \item For $\tau$ decaying to electron with a branching ratio of 17.4\%, we assume a signal efficiency of 30\% considering $\nu_e$ events as possible background. We consider the signal to background ratio of 2.45 in our analysis~\cite{Ghoshal:2019pab}.
    
\end{itemize}
The muonic decays have not been taken into account since the discrimination of the number of background events would be too large compared to the signal events~\cite{DeGouvea:2019kea,Ghoshal:2019pab}. The total number of $\nu_\tau$ ($\bar{\nu}_\tau$) events in DUNE is expected to be roughly 72 (37) per year.
The normalization error for the signal is taken to be 20\%. The results of our analysis for $\alpha_{33}$ and $|\alpha_{32}|$ are shown in Fig.~\ref{fig:tau} and Table~\ref{tab:DUNE_taus}. The allowed range of $\alpha_{33}$, which appears only at the second order in the probability, is reduced of $\sim$ 13\% by the inclusion of the new oscillation channel, and the new limits are set into the range [-0.16, 0.15] (see Table~\ref{tab:DUNE_taus}). On the other hand, the sensitivity to $|\alpha_{32}|$, which impacts linearly the $\nu_\tau$ appearance probability, is significantly improved: in this case, the new upper bound is roughly 60\% smaller than the one set by the standard oscillation channels, namely $|\alpha_{32}|<0.19$. 

\section{Summary and conclusions}
\label{sec:conclusion}
Neutrino oscillation has been one of the most important discoveries over the last few decades and a large number of pioneering neutrino experiments have confirmed this phenomena of mass-induced flavor transition. 
Excellent data from various neutrino oscillation experiments having different baselines and energies have enabled us to achieve remarkable precision on almost all the three-flavor oscillation parameters.
 However, some of the features of neutrino mass-mixing parameters, like, neutrino mass hierarchy, the value of the CP-violating phase $\delta_{\mathrm{CP}}$, and octant of $\theta_{23}$ are still poorly known and next-generation neutrino experiments will play an important role to address these issues with high confidence level.
 
Given the magnificent precision on neutrino mixing angles and rapidly increasing knowledge on $\delta_{\mathrm{CP}}$, 
it seems quite natural to ask if there is any violation of the unitary property of $3\times 3$ PMNS mixing matrix, which may be related to the existence of new mass eigenstates of neutrinos.
In this context, there exists one well-known parameterization in the literature, which takes into account the non-unitary neutrino mixing (NUNM) by introducing a lower triangular matrix with three real and three complex parameters, $\alpha_{ij}$.

To study the impact of these NUNM parameters on various oscillation channels, for the first time, we derive simple 
approximate analytical expressions for the oscillation probabilities in matter in the atmospheric regime ($\Delta_{31}>>\Delta_{21}$).
Our perturbative expansions are valid up to second order in the small deviations from the mixing angles by their tri-bimaximal values, second order in the NUNM parameters, and first order in the matter potential. We have shown that the $\nu_{\mu}\rightarrow\nu_e$ appearance probability (see Eq.~\ref{eq:pme_anylit_main}) mainly depends on $|\alpha_{21}|$, but when matter potential is large, the impact of $|\alpha_{31}|$ and $|\alpha_{32}|$ can also be significant. On the other hand, the $\nu_{\mu}\rightarrow\nu_{\mu}$ disappearance probability (see Eq.~\ref{eq:pmm_anylit_main}) mainly relies on $\alpha_{22}$, but sub-leading dependencies on $|\alpha_{21}|$ in vacuum and $\alpha_{33}$, $|\alpha_{31}|$, and $|\alpha_{32}|$ in matter are also present. The only parameter that does not appear in our formulae is $\alpha_{11}$, which is only relevant at the higher-orders in our perturbative expansions  as shown in Ref.~\cite{Escrihuela:2015wra} for the vacuum case.

In this work, we analyze in detail, the impact of possible NUNM in the context of long-baseline experiments DUNE and T2HK having one detector in Japan (JD) and a second detector in Korea (KD), and the combination of these two detectors, popularly known as T2HKK or JD+KD. First, we show how the  various NUNM parameters ($\alpha_{ij}$) are correlated with the oscillation parameters $\theta_{23}$ and $\delta_{\mathrm{CP}}$ for these setups.
Then, we estimate in detail the sensitivities of these experiments to place direct, model-independent, competitive constraints on the six NUNM parameters  at 90\% C.L. (see Fig.~\ref{fig:chi_sq} and Table~\ref{Tab:constraints}). 
 The wide-band neutrino beam in DUNE encompassing both first and second oscillation maxima allows us to measure the NUNM parameters at several $L/E$ values in the presence of significant matter effect due to its 1300 km baseline. Indeed, DUNE 
 shows better sensitivity than JD+KD in constraining the NUNM parameters, which are influenced by matter effects, namely, $\alpha_{33}$, $|\alpha_{31}|$, and $|\alpha_{32}|$.
We observe that JD+KD will provide a stringent constraint on $\alpha_{22}$ as compared to DUNE because it has less systematic uncertainties in the disappearance channel. 
Also, due to the larger statistics in the appearance channel, JD+KD will give significantly better limit on the NUNM parameter $|\alpha_{21}|$ in comparison to DUNE.
Lastly, $\alpha_{11}$ is expected to be constrained almost in the same way by these two experiments.
We show how the limits on the NUNM parameters get improved in case of DUNE, if the total exposure is increased from 336 kt-MW-yrs to 480 kt-MW-yrs (corresponding to an increase in the total run-time  from 7 years to 10 years), as proposed in the recent TDR~\cite{DUNE:2021cuw}.
 We also estimate how much the sensitivities can be improved by adding the prospective data from DUNE and JD+KD. Finally, we compare our results with the constraints that can be achieved using the full exposure of currently running experiments T2K and NO$\nu$A.
 
 Due to the so-called zero-distance effects which are induced by the non-unitary neutrino mixing in neutrino oscillation probabilities,
the prospective data from near detectors in both DUNE and JD+KD experiments could be in principle used to bound the three NUNM parameters $|\alpha_{21}|$, $\alpha_{11}$ and $\alpha_{22}$. However, the zero-distance effect should also be taken into account if the near detectors data will be used to measure the initial neutrino flux for both experiments. This would lead to a substantial deterioration of the  limits that the far detectors could set on $\alpha_{11}$ and $\alpha_{22}$, as summarized in Table~\ref{Table:cons_w_norm}.

Moreover, in DUNE, the expected limits on $|\alpha_{32}|$ and $\alpha_{33}$ get improved by ${\cal O}(20)\%$ when we also add the $\nu_{\tau}$ appearance sample in our analysis. 

Long-baseline experiments have contributed significantly in our journey to establish the three-neutrino paradigm on a strong footing. Future high-precision long-baseline experiments such as DUNE and T2HKK (JD+KD) will continue this legacy and rigorously test the unitarity of the PMNS matrix by measuring the mixing angles and CP phase $\delta_{\mathrm{CP}}$ with remarkable precision. We hope our present work will provide further boost along this direction.

\subsection*{Acknowledgments}
 We would like to thank M. Blennow, E. Fernandez Martinez, S. Roy, S. Sahoo, M. Singh, and M. Tortola for useful discussions.
 S.K.A. is supported by the INSPIRE
Faculty Research Grant [IFA-PH-12] from the Department
of Science and Technology (DST), Govt. of India.
S.K.A. acknowledges the financial support from the
Swarnajayanti Fellowship Research Grant
(No. DST/SJF/PSA-05/2019-20) provided by
the DST, Govt. of India and the Research Grant
(File no. SB/SJF/2020-21/21) from the Science
and Engineering Research Board (SERB) under
the Swarnajayanti Fellowship by the DST, Govt. of India. S.K.A would like to thank the United States-India Educational Foundation for providing the financial support through the Fulbright-Nehru Academic and Professional Excellence Fellowship (Award No. 2710/F-N APE/2021).
The numerical simulations are carried out using SAMKHYA: High-Performance Computing Facility at Institute of Physics, Bhubaneswar.

\begin{appendix}

\renewcommand\thefigure{A\arabic{figure}}
\renewcommand\theequation{A\arabic{equation}}
\setcounter{figure}{0} 
\setcounter{equation}{0}

\section{Analytical expressions of the transition probabilities in the regime of small matter effects}
\label{Appendix:A}
\subsection*{Exact vs. approximate probabilities}
\begin{figure}[h!]
\centering
\includegraphics[width=\textwidth]{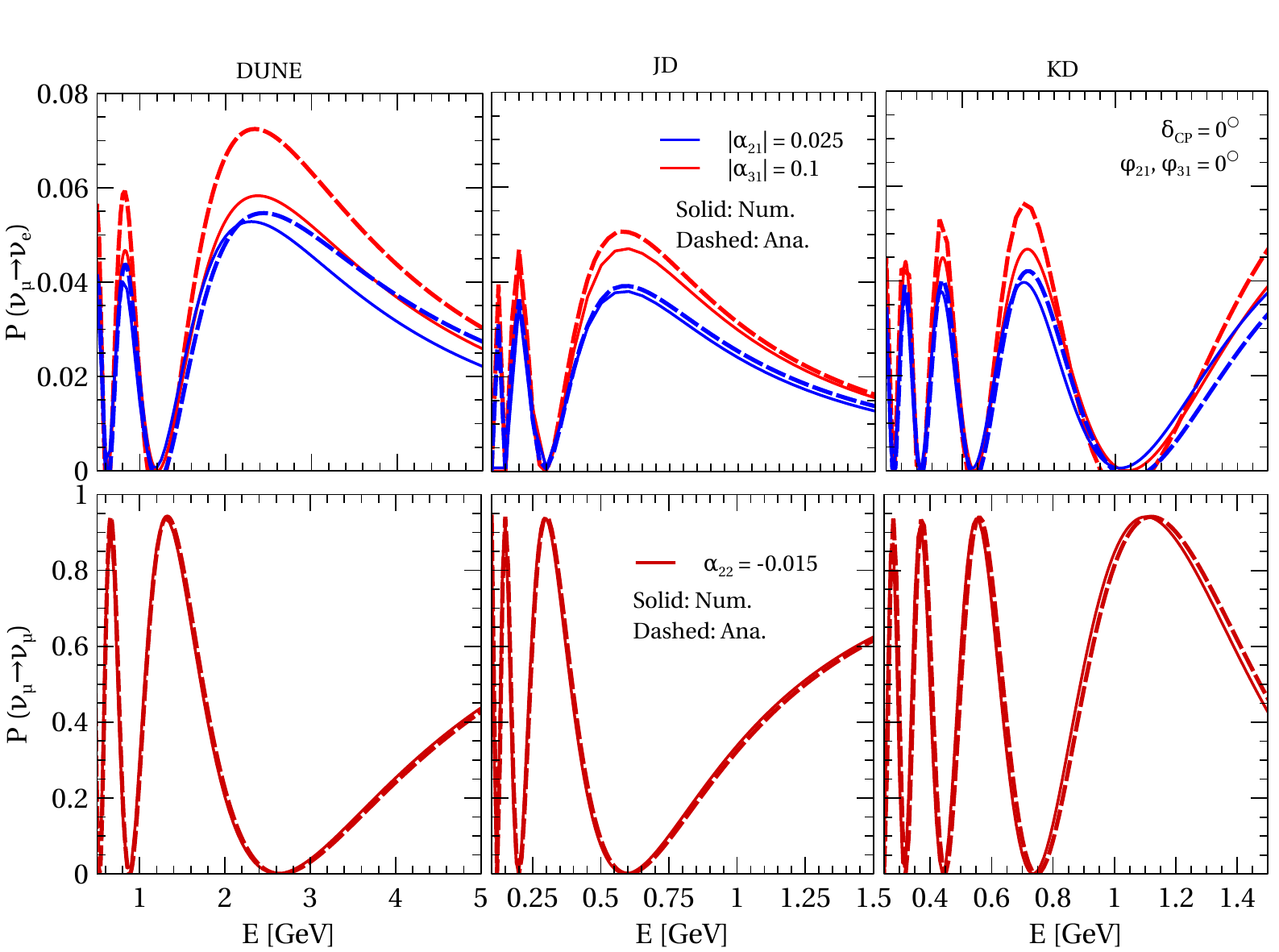}
\mycaption{Upper panels show the $\nu_{\mu}\rightarrow\nu_{e}$ appearance probability in presence of the NUNM parameters $|\alpha_{21}|\,=\,0.025$ and $|\alpha_{31}|\,=\,0.1$. Lower panels depict the $\nu_{\mu}\rightarrow\nu_{\mu}$
disappearance probability for $\alpha_{22}\,=\,0.015$. 	
Left, middle, and right columns correspond to the baselines of 1300 km (DUNE), 295 km (JD), and 1100 km (KD), respectively.
Solid curves in each panel show the oscillation probabilities obtained numerically, while the dashed curves are the probabilities obtained analytically using Eq.~\ref{eq:pme_anylit_main} and Eq.~\ref{eq:pmm_anylit_main}. We consider the values of  $\delta_{\mathrm{CP}}$ and the phases $\phi_{21}$, $\phi_{31}$ equal to zero. The values of the other standard oscillation parameters are taken from Table~\ref{table:vac}. }
\label{fig:prob_mueCheck}
\end{figure}

In this appendix, we provide a check of the goodness of our expanded probabilities,  Eqs.~\ref{eq:pme_anylit_main} to \ref{eq:pmm_anylit_main}. To this aim, in the upper panels of Fig.~\ref{fig:prob_mueCheck}, we show the $\nu_{\mu}\rightarrow\nu_{e}$  appearance probabilities as a function of neutrino energy considering the NUNM parameters 
$|\alpha_{21}|$ and $|\alpha_{31}|$ one at a time. We portray the same in the lower panels for $\nu_{\mu}\rightarrow\nu_{\mu}$  disappearance channel in presence of the diagonal NUNM parameter $\alpha_{22}$.
We consider three different baselines, corresponding to DUNE (left panels), JD (middle panels), and KD (right panels). In each panel, solid curves are obtained numerically from the full (not expanded) Hamiltonian, while dashed curves correspond to the oscillation probabilities obtained analytically using Eq.~\ref{eq:pme_anylit_main} and Eq.~\ref{eq:pmm_anylit_main}. The value of the CP phase $\delta_{\mathrm{CP}}$ and phases $\phi_{21}$, $\phi_{31}$ are fixed to zero, while the values of the other standard oscillation parameters are taken from Table\ \ref{table:vac}. As we can see, the agreement between the full solid and the analytical dashed curves is generally very good for both probabilities.

\subsection*{Perturbative expansion of $P_{ee}$, $P_{e\tau}$, and $P_{\mu\tau}$}
For the sake of completeness,  we give here the perturbative expressions of  the probabilities $P_{ee}$, $P_{e\tau}$ and $P_{\mu\tau}$, using the same approximations discussed in Sec.~\ref{formulae}. We start with the $\nu_e\rightarrow\nu_e$ disappearance channel: 
\begin{eqnarray}
P_{e e } &=& 1 + 4 \alpha_{11} +6 \alpha_{11}^2-\left(\frac{4 r }{\Delta_{31}}\right)\Delta_n \left[|\alpha_{21}| \cos(\delta-\phi_{21})+ \alpha_{31} \cos(\delta-\phi_{31})\right]\sin^2\Delta_{31} +\nonumber \\
&& -\left(\frac{2 r }{\Delta_{31}}\right) \sin\Delta_{31}\left[(\Delta_{31}+2\Delta_e)\sin\Delta_{31}-2\Delta_{31}\Delta_e \cos\Delta_{31}\right] \,.
\label{eq:e-disapp}
\end{eqnarray}
The $\nue\rightarrow\nu_\tau$ transition is governed by the following expression: 
\begin{eqnarray}
P_{e \tau} &=&  \left(\frac{r^2}{\Delta_{31}}\right) \sin \Delta_{13} \left[(\Delta_{31}+2 \Delta_e) \sin \Delta_{31}-2 \Delta_{31} \Delta_e \cos\Delta_{31}\right]\nonumber \\
&& \left(\frac{|\alpha_{31}|^2}{\Delta_{31}}\right)\left[\Delta_{31}-\Delta_n (1-\cos2\Delta_{31})\right]+
\nonumber \\
&& \left(\frac{2 |\alpha_{21}|  r }{\Delta_{31}}\right)\Delta_n\sin\Delta_{31} \left[\sin\Delta_{31} \cos (\delta_{\mathrm{CP}}-\phi_{21})-\Delta_{31} \cos
   (\delta_{\mathrm{CP}}-\Delta_{31}-\phi_{21})\right]+
 \\
&&\left(\frac{|\alpha_{31}| r }{\Delta_{31}}\right)\left[2 \Delta_{31} \sin\Delta_{31} (\Delta_e+\Delta_n) \cos (\delta_{\mathrm{CP}}-\Delta_{13}-\phi_{31})+\right. \nonumber \\
&&\left.\sin (\delta_{\mathrm{CP}}-\Delta_{31}-\phi_{31}) (\sin\Delta_{31} (2 \Delta_{31}+2\Delta_e-\Delta_n)-2 \Delta_{31} \Delta_e \cos \Delta_{31})+\right. \nonumber \\ && \nonumber \left. \Delta_n \sin\Delta_{31} \sin (\delta
+\Delta_{31}-\phi_{31})\right]+\nonumber\\
&&
\left(\frac{|\alpha_{21}| |\alpha_{31}|}{\Delta_{31}}\right) \Delta_n \left[-2 \Delta_{31} \sin (\phi_{21}-\phi_{31})+\cos (2 \Delta_{31}-\phi_{21}+\phi_{31})-\cos (\phi_{21}-\phi_{31})\right]\,.\nonumber
\end{eqnarray}
Eventually, the $\nu_{\tau}$ appearance from a $\nu_{\mu}$ beam is regulated by the following probability expression:
\begin{eqnarray}
P_{\mu \tau} &=& \sin^2\Delta_{31}\left(1+2 \alpha_{22}+2 \alpha_{33}-4 a^2+ \alpha_{22}^2+ \alpha_{33}^2 + 4 \alpha_{22} \alpha_{33}\right)  +\nonumber \\
&&|\alpha_{32}| \sin2\Delta_{13} \left[2 \Delta_n   \cos \phi _{32}-\sin\phi_{32}\right] +\nonumber \\
&& \left(\frac{r^2}{\Delta_{31}}\right) \sin (\Delta_{31}) \left[2 \Delta_{31} \Delta_e\cos\Delta_{31}-\sin\Delta_{31} (\Delta_{31}+2 \Delta_e )\right] + \nonumber \\
&& \left(\frac{2 }{\Delta_{31}}\right)\left(|\alpha_{21}|^2+|\alpha_{31}|^2\right)\Delta_{n} \sin\Delta_{31} \left[\sin\Delta_{31}-\Delta_{31} \cos\Delta_{31}\right] + \nonumber \\
&&\left(\frac{|\alpha_{32}|^2}{\Delta_{31}}\right) \left[\Delta_n  \sin2 \phi_{32} (\sin2 \Delta_{31}-2 \Delta_{31} \cos2 \Delta_{31})+\Delta_{31} \cos ^2\Delta_{13}\right] +  \nonumber \\
&&\left(\frac{2 |\alpha_{21}| r} {\Delta_{31}}\right)\sin \Delta_{31} \left[\sin\Delta_{31} \cos (\delta_{\mathrm{CP}}-\phi_{21}) (\Delta_{31}+\Delta_e -\Delta_n )-\Delta_{31} \Delta_e   \cos (\delta_{\mathrm{CP}}+\Delta_{31}-\phi_{21})+\right.\nonumber \\
&& \qquad \qquad \left. \Delta_{31} \Delta_n \cos (\delta_{\mathrm{CP}}-\Delta_{31}-\phi _{21})\right] + \nonumber \\
&&\left(\frac{2 |\alpha_{31}| r}{\Delta_{31}}\right) \left[\sin \Delta_{31}(\sin\Delta_{31} \cos(\delta_{\mathrm{CP}}-\phi_{31}) (\Delta_{31}+\Delta_e -\Delta_n )-\Delta_{31} \Delta_e  \cos (\delta_{\mathrm{CP}}-\Delta_{31}-\phi_{31})+ \right.\nonumber \\ &&  \left. \Delta_{31} \Delta_n   \cos (\delta_{\mathrm{CP}}+\Delta_{31}-\phi_{31})\right] + \left(\frac{8 a}{\Delta_{31}}\right)\left(\alpha_{22}-\alpha_{33}\right) \Delta_n \sin\Delta_{31} (\Delta_{31} \cos\Delta_{31}-\sin \Delta_{31}) + \nonumber  \\
&& 4 a |\alpha_{32}| \sin^2\Delta_{31} \cos\phi_{32} +\nonumber \\
&& \left(\frac{2 |\alpha_{21}| |\alpha_{31}|}{\Delta_{31}}\right) \sin\Delta_{31} \left[2 \Delta_{31} (\Delta_e+\Delta_n) \cos (\Delta_{31}-\phi_{21}+\phi_{31})+2 \Delta_n 
\sin\Delta_{31} \cos (\phi_{21}-\phi_{31})-\right.\nonumber \\
&&  \left.\Delta_{31} \sin (\Delta_{31}-\phi_{21}+\phi_{31})\right] + \nonumber \\ 
&&4 \alpha_{22} |\alpha_{32}| \sin\Delta_{31} \left(\frac{\Delta_n   \cos\phi_{32} (\sin\Delta_{31}+2 \Delta_{31} \cos\Delta_{31})}{\Delta_{31}}-\cos \Delta_{31} \sin\phi_{32}\right) + \nonumber \\
&& -\left(\frac{2 |\alpha_{32}| \alpha_{33}}{\Delta_{31}}\right) \sin\Delta_{31} \left[2 \Delta_{n} \cos\phi_{32} (\sin\Delta_{31}-3 \Delta_{31} \cos \Delta_{31})+\Delta_{31} \cos\Delta_{31}\sin\phi _{32}\right]\,.
\label{tauprob}
\end{eqnarray}
The zero-distance effects are as follows:
\begin{eqnarray}
	\label{eq:zero-distance}
P_{e e}^{L = 0} &\sim& 1 + 4 \alpha_{11} + 6 \alpha_{11}^2 \,, \nonumber \\
P_{e\tau}^{L = 0} &\sim&  |\alpha_{31}|^2\,,\\
P_{\mu\tau}^{L = 0} &\sim&  |\alpha_{32}|^2\,.\nonumber 
\end{eqnarray}
It is easy to check that, when all $\alpha$ parameters are set to zero, we recover the unitary relation $P_{\mu e}+P_{\mu \mu}+P_{\mu \tau}=1$.


\renewcommand\thefigure{B\arabic{figure}}
\renewcommand\thetable{B\arabic{table}}
\renewcommand\theequation{B\arabic{equation}}
\setcounter{figure}{0} 
\setcounter{table}{0}
\setcounter{equation}{0}

\section{Impact of marginalization over the  other NUNM parameters while showing the constraints on $\alpha_{ij}$}
\label{sec:NU_marg}

 \begin{figure*}[h!]
 	\centering
 	\includegraphics[width=\textwidth]{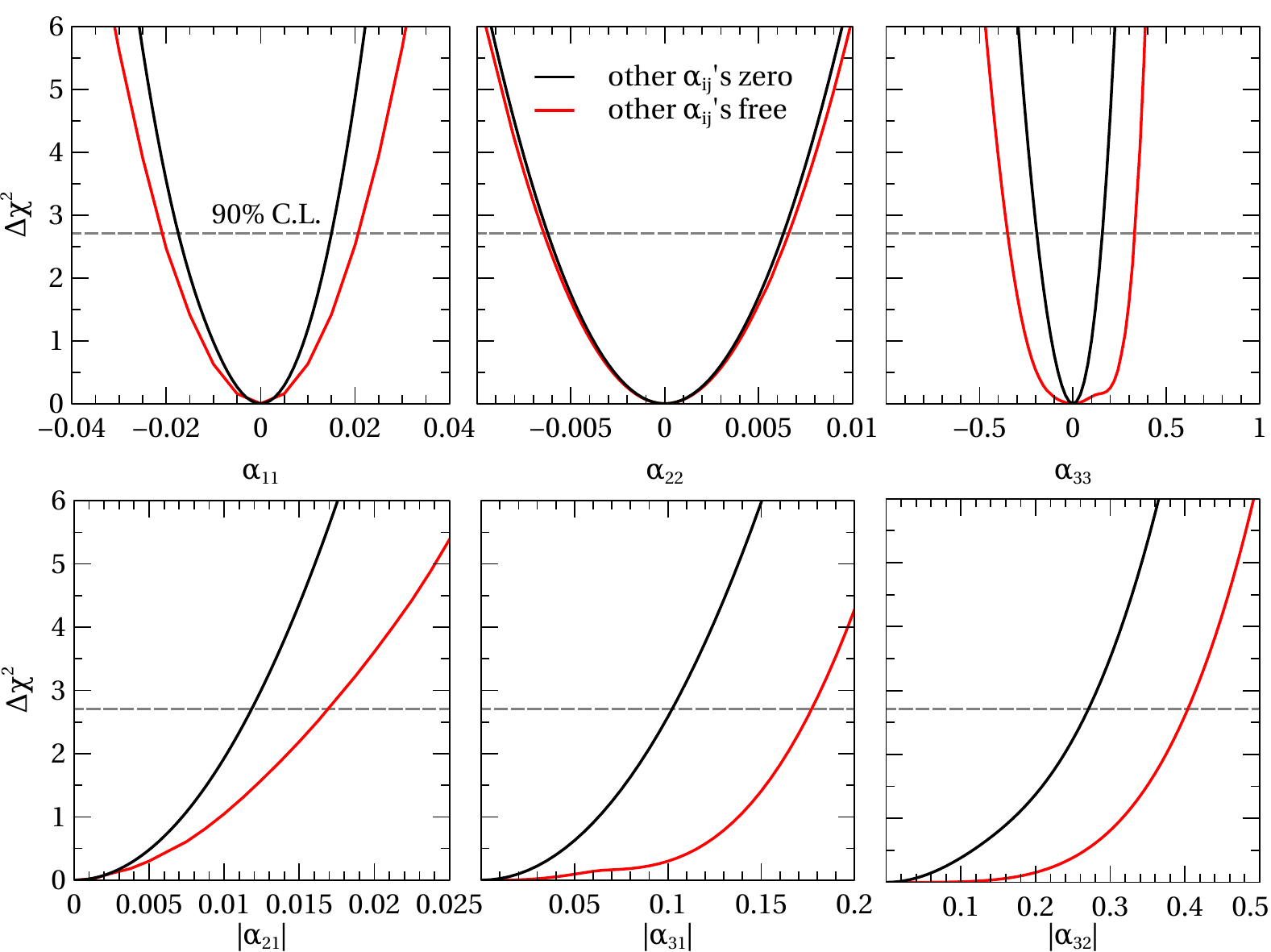}
 	\mycaption{The combined sensitivity of DUNE and JD+KD setups to the NUNM parameters. True values of the standard oscillation parameters are given in Table~\ref{table:vac}. We marginalize over $\theta_{23}$ and $\delta_{\mathrm{CP}}$ in the fit (see text for details). Black curves show the sensitivity when only one NUNM parameter is considered at a time, while the others are taken to be zero. Red curves in all the panels correspond to the case when all other NUNM parameters are kept free in the fit.  \label{fig:NU-param-marg}} 
 \end{figure*}
 
In this appendix, we derive the constraints on the NUNM parameters obtained, differently from the procedure adopted in the main text, marginalizing over the other NUNM parameters $\alpha_{ij}$. Our numerical results are shown in Fig.~\ref{fig:NU-param-marg}, where we show the impact of the marginalization over various NUNM parameters on the sensitivity of the DUNE+JD+KD setup. 
 In each panel, black curves show the sensitivity for a particular NUNM parameter, when all other NUNM parameters are fixed in the fit. The red curves show the situation when all un-displayed NUNM parameters are marginalized in the fit with no priors.
The 90\% C.L. (1 d.o.f.) constraints on various NUNM parameters are summarized in Table~\ref{tab:NUNM_marg}.
 \begin{table}
 	\centering
 	\begin{tabular}{|c|c|c|}
 		\hline\hline
 		Parameter & Other $\alpha_{ij}$'s zero & Other $\alpha_{ij}$'s free \\ 
 		\hline
 		$\alpha_{11}$& [-0.017, 0.017] & [-0.02, 0.02] \\
 		\hline
 		$\alpha_{22}$ & [-0.006, 0.006] &  [-0.006, 0.006] \\
 		\hline
 		$\alpha_{33}$ & [-0.20, 0.17] & [-0.35, 0.33] \\
 		\hline
 		$|\alpha_{21}|$ & $<$ 0.012  & $<$ 0.017\\ 
 		\hline
 		$|\alpha_{31}|$ & $<$ 0.11 & $<$ 0.18\\
 		\hline
 		$|\alpha_{32}|$ & $<$ 0.27 & $<$ 0.40\\
 		\hline\hline
 	\end{tabular}
 	\mycaption{90\% C.L.(1 d.o.f.) limits on various NUNM parameters. Second column shows the constraints considering only one NUNM parameter at a time, while other NUNM parameters are assumed to be zero in the fit. Third column depicts the bounds on a given NUNM parameter when all other NUNM parameters and the phases associated with the off-diagonal parameters are kept free in the fit. True values of the standard oscillation parameters are taken from Table~\ref{table:vac}. We marginalize over $\theta_{23}$ and $\delta_{\mathrm{CP}}$ in the fit (see text for details). \label{tab:NUNM_marg}} 
 \end{table}
We observe that the marginalization over all other NUNM parameters worsen the bounds on all of them but $\alpha_{22}$. This is due to the fact that, as it can be seen in Eq. \ref{eq:pmm_anylit_main}, $\alpha_{22}$ appears at the leading order, with no correlations to the other $\alpha_{ij}$.
The other diagonal parameter $\alpha_{11}$ shows only a marginal deterioration at the level of 15\% since, as before, its correlation with the other NUNM parameters are mild in $\nu_e$ appearance probability as shown in Ref.~\cite{Escrihuela:2015wra}.
We see from Table~\ref{tab:NUNM_marg} that there is a considerable deterioration in the sensitivities for $\alpha_{33}$, $|\alpha_{21}|$, $|\alpha_{31}|$, and $|\alpha_{32}|$ when we marginalize over the un-displayed NUNM parameters in the fit.  
This happens because all these four parameters are strongly correlated among them and with other two NUNM parameters $\alpha_{11}$ and $\alpha_{22}$ (see Eqs.~\ref{eq:pme_anylit_main} and \ref{eq:pmm_anylit_main}).
Note that $|\alpha_{21}|$ does not have any correlation with the other NUNM parameters in $\nu_\mu\rightarrow\nu_e$ appearance channel but it is strongly correlated with $\alpha_{22}$, $|\alpha_{31}|$, and $|\alpha_{32}|$ in $\nu_\mu\rightarrow\nu_\mu$ disappearance channel which reduces its sensitivity by roughly 40\%, when we keep the other NUNM parameter free in the fit.
Similar correlations among the NUNM parameters are also responsible for worsening the bounds on $|\alpha_{31}|$ by 60\%, $|\alpha_{32}|$ by 45\%, and by a factor of two for $\alpha_{33}$ when we marginalize over the other NUNM parameters in the fit.

\end{appendix}
\newpage
\bibliographystyle{JHEP}
\bibliography{reference_NU}
\end{document}